\documentclass[twocolumn]{aastex631}
\usepackage{amsmath} 
\usepackage{appendix}

\usepackage{CJK}
\usepackage{upgreek}
\usepackage{booktabs}

\begin{document}

\title{Propagation-induced Frequency-dependent Polarization Properties of Fast Radio Burst\footnote{Released on March, 1st, 2021}}
\begin{CJK*}{UTF8}{gbsn}

\correspondingauthor{Wei-Yang Wang}
\email{wywang@ucas.ac.cn}

\author[0000-0001-9036-8543]{Wei-Yang Wang (王维扬)}
\affiliation{School of Astronomy and Space Science, University of Chinese Academy of Sciences, Beijing 100049, Peopleʼs Republic of China} 

\author[0000-0002-2552-7277]{Xiaohui Liu (刘小辉)}
\affiliation{National Astronomical Observatories, Chinese Academy of Sciences, Beijing 100101, Peopleʼs Republic of China}
\affiliation{School of Astronomy and Space Science, University of Chinese Academy of Sciences, Beijing 100049, Peopleʼs Republic of China} 

\author[0000-0001-7931-0607]{Dongzi Li (李冬子)}
\affiliation{TAPIR, California Institute of Technology, Pasadena, CA 91125, USA}
\affiliation{Department of Astrophysical Sciences, Princeton University, Princeton, NJ 08544, USA}

\author[0000-0002-9725-2524]{Bing Zhang (张冰)}
\affiliation{The Nevada Center for Astrophysics, University of Nevada, Las Vegas, NV 89154}
\affiliation{Department of Physics and Astronomy, University of Nevada, Las Vegas, NV 89154, USA}

\author[0000-0001-6651-7799]{Chen-Hui Niu (牛晨辉)}
\affiliation{Institute of Astrophysics, Central China Normal University, Wuhan 430079, Peopleʼs Republic of China}

\author{Jifeng Liu (刘继峰)}
\affiliation{School of Astronomy and Space Science, University of Chinese Academy of Sciences, Beijing 100049, Peopleʼs Republic of China} 
\affiliation{New Cornerstone Science Laboratory, National Astronomical Observatories, Chinese Academy of Sciences, Beijing 100101, Peopleʼs Republic of China}
\affiliation{Institute for Frontiers in Astronomy and Astrophysics, Beijing Normal University, Beijing 102206, Peopleʼs Republic of China}

\author[0000-0002-9042-3044]{Renxin Xu (徐仁新)}
\affiliation{School of Physics and State Key Laboratory of Nuclear Physics and Technology, Peking University, Beijing 100871, Peopleʼs Republic of China}
\affiliation{Kavli Institute for Astronomy and Astrophysics, Peking University, Beijing 100871, Peopleʼs Republic of China}

\author[0000-0001-5105-4058]{Weiwei Zhu (朱炜玮)}
\affiliation{National Astronomical Observatories, Chinese Academy of Sciences, Beijing 100101, Peopleʼs Republic of China}
\affiliation{Institute for Frontiers in Astronomy and Astrophysics, Beijing Normal University, Beijing 102206, Peopleʼs Republic of China}

\author{Kejia Lee (李柯伽)} 
\affiliation{National Astronomical Observatories, Chinese Academy of Sciences, 20A Datun Road, Chaoyang District, Beijing 100101, People's Republic of China}
\affiliation{Kavli Institute for Astronomy and Astrophysics, Peking University, Beijing 100871, People's Republic of China}
\affiliation{Department of Astronomy, Peking University, Beijing 100871, People's Republic of China}
\begin{abstract}

Frequency-dependent polarization properties provide crucial insights into the radiation mechanisms and magnetic environments of fast radio bursts (FRBs).
We explore an analytical solution of radiative transfer of the polarization properties of FRBs as a strong incoming wave propagates in a homogeneous magnetized plasma. The cases of a thermal plasma is studied in detail. 
The rotational axis of the polarization spectrum undergoes precession with frequency on the Poincar\'e sphere when the medium has both strong Faraday rotation and conversion.
Such precession on the Poincar\'e sphere could occur in hot or cold plasma with a strong magnetic field component perpendicular to the line of sight.
Significant absorption can exist in a dense plasma medium, which may give rise to a highly circularly polarized outgoing wave.
We apply the analytical solution with the mixing Faraday case to fit the observations of frequency-dependent Stokes parameters for FRB 20180301A and FRB 20201124A.
The analytical solution offers a more physical description of FRBs' magnetic environment properties than the empirical ``generalized Faraday rotation'' method commonly adopted in the literature.
The frequency-dependent Stokes parameters may be associated with reversing rotation measures or the presence of a persistent radio source around an FRB.
\end{abstract}



\section{Introduction} \label{sec:intro}
Fast radio bursts (FRBs) are extragalactic bursts lasting just milliseconds \citep{lorimer07,thornton13,petroff19}, yet their origins remain a mystery \citep{zhang23}.
Among the hundreds and thousands of {FRB sources detected} \footnote{See Transient Name Server, https://www.wis-tns.org/.}, one has been detected within the Milky Way, whose source was confirmed as a Galactic magnetar \citep{2020Natur.587...59B,2020Natur.587...54C}. This suggests that at least some FRBs originate from magnetars.

Polarization features of such individual bright pulses have garnered significant attention due to their potential insights into the underlying astrophysical processes.
Although only a small fraction of FRB sources have been measured for polarization, they exhibit various polarization properties.
Judging solely by the polarization properties, repeating and apparently non-repeating FRBs seem to lack marked differences. Most FRBs display flat polarization position angles (PAs) \citep{2018Natur.553..182M,2021ApJ...908L..10H,2021NatAs...5..594N}, while others show regular or irregular PA variations \citep{2020Natur.586..693L,2022RAA....22l4003J,2022MNRAS.512.3400K,2022Natur.611E..12X,2025Natur.637...43M}.
Notably, three bursts from FRB 20201124A exhibit sudden jumps in PA \citep{2024ApJ...972L..20N}.
Most bursts are highly linearly polarized, some show significant circular polarization \citep{2020MNRAS.497.3335D,2022MNRAS.512.3400K,2022Natur.611E..12X,2023ApJ...955..142Z,2024arXiv240803313J}.
These polarization properties might reflect intrinsic mechanisms arising from the magnetosphere of neutron stars \citep{2022MNRAS.517.5080W,2022ApJ...927..105W,2022ApJ...925...53Z,2023MNRAS.522.2448Q,2024ApJ...972..124Q,2024arXiv240804401Z}.

Some polarization properties are readily attributed to propagation effects. The most straightforward one is Faraday rotation, which is caused by the propagation delay between left- and right-handed circular polarization modes when a linearly polarized radio wave travels along a magnetic field through a cold magnetized plasma. This has been quantitatively described by the rotation measure (RM) that has been measured in many FRB sources. Some FRB sources show long-term RM evolution; some show RM variations within a timescale as short as several days or even hours \citep{2018Natur.553..182M,2020Natur.586..693L,2022Natur.611E..12X,2024arXiv241010172X}; while some others such as FRB 20220912A show a near-zero and stable RM over two months \citep{2023ApJ...955..142Z,Feng_2024}. A good fraction of repeating FRB sources, e.g. FRB 20180301A, FRB 20190303A, FRB 20190520B, and FRB 20200929C,  were reported to show sign changes of RM, which is indicative of reversal of the parallel component of the magnetic field configuration along the line of sight \citep{2023Sci...380..599A,2023MNRAS.526.3652K,ng2024polarizationproperties28repeating}.

Frequency-dependent polarization properties, including linear and circular polarization degrees as well as PAs, have been identified in a subset of bursts from FRB 20180301A, FRB 20201124A, FRB 20210912A, and FRB 20230708A \citep{2022MNRAS.512.3400K,2022Natur.611E..12X,2023MNRAS.526.3652K,2024arXiv241114784B,2024MNRAS.534.2485U}.
Such phenomena have also been discovered in two radio pulsars, the magnetar XTE J1810--197 and the long-period radio transient GPM J1839--10 \citep{2024NatAs...8..606L,doi:10.1126/sciadv.adp6351}.
Their frequency spectra of both circular and linear polarization intensities oscillate with wavelength or frequency, which have been empirically interpreted as the signature of ``generalized Faraday rotation'' (GFR) \citep{1997JPlPh..58..735M,1998PASA...15..211K}.
This effect involves the partial conversion of linear polarization into circular polarization, driven by the process of Faraday conversion.
However, the empirical GFR modeling approach assumes a power-law dependence of the Stokes parameters on frequency and fits the geometric parameters of the trajectory on the Poincar\'e sphere, which only provides a phenomenological description of the observations and fails to reflect the true physical information of the medium.

These key observational results call for developing a physical propagation model of polarization properties of radio waves, which can be used to diagnose the environments of FRBs. A realistic plasma medium around an FRB source likely mixes Faraday rotation and conversion with possible accompanying radiation and absorption (e.g. from a persistent radio source). There is no catch-all model for such a complex environment to describe
the observational effects using a simple model involving standard Faraday rotation. 
We explore a more general model of radiative transfer of polarization properties of radio waves in a complex environment involving Faraday rotation, conversion, and absorption and derive analytical solutions. We apply the model to discuss some complex frequency-dependent polarization properties observed in some FRBs. The paper is organized as follows. The radiative transfer of the polarized emission is discussed in two general scenarios in Section \ref{sec2}.
We exhibit some spectro-polarimetric simulation results in Section \ref{sec3}.
Some implications are discussed in Sections \ref{sec4} and \ref{sec5}, and the consequences are summarized in Section \ref{sec6}.

\end{CJK*}

\section{Radiative transfer of polarized emission in a magnetized plasma medium} \label{sec2}
In Section \ref{sec2.3}, we provide an analytical solution to the radiative transfer equation of polarized waves. The elements of the matrix in the transformation equation, for instance, emissivity, absorption and Faraday mixing coefficients, are determined by the energy distribution of the medium plasma.
Meanwhile, the relativistic effects of the plasma also affect the propagation of FRB waves in the medium.
We discuss both cold (non-relativistic) and hot (relativistic) plasma in Section \ref{sec2.2}.

\subsection{Radiative transfer of polarized waves}\label{sec2.3}
\begin{figure}
\centering
\includegraphics[width=0.5\textwidth]{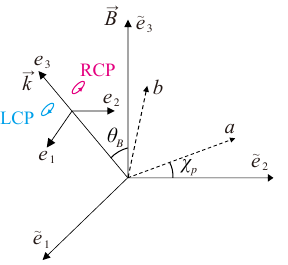}
\caption{Sketch map of polarization: vectors $e_1$, and $e_2$ represent the wave vector, respectively. $e_1$ is located in the $k-B$ plane and $e_2$ is perpendicular to $e_1$ and parallel to $\Tilde{e}_2$. In the coordinate, a left-handed circular polarization (LCP) is shown in cyan and magenta for right-handed circular polarization (RCP). $\chi_p$ and $d\chi_p$ represent the electric vector position angle defined in the arbitrary coordinate system $(\boldsymbol{a}\,,\boldsymbol{b})$ and the change of it due to the radiation transfer.}
\label{fig:e1e2e3}
\end{figure}

Consider a polarized radiation propagating in a homogeneous magnetized plasma with complex properties. The transfer of the polarized emission in the Stokes basis can be described by \citep{1969SvA....13..396S}
\begin{equation}
\frac{d\vec{S}}{ds}=\vec{\epsilon}-\boldsymbol{M}\vec{S},
\end{equation}
or explicitly
\begin{equation}
\frac{d}{d s}\left(\begin{array}{l}
I \\
Q \\
U \\
V
\end{array}\right)=\left(\begin{array}{l}
\epsilon_I \\
\epsilon_Q \\
\epsilon_U \\
\epsilon_V
\end{array}\right)-\left(\begin{array}{cccc}
\eta_I & \eta_Q & \eta_U & \eta_V \\
\eta_Q & \eta_I & \rho_V & -\rho_U \\
\eta_U & -\rho_V & \eta_I & \rho_Q \\
\eta_V & \rho_U & -\rho_Q & \eta_I
\end{array}\right)\left(\begin{array}{l}
I \\
Q \\
U \\
V
\end{array}\right) ,
\label{eq:trans1}
\end{equation}
where $s$ is the path length, $\epsilon_S$, $\eta_S$ and $\rho_S$ are emissivity, absorption and Faraday mixing coefficients, respectively.
Note that by rotating the coordinate system, $U=0$ can be satisfied for a single electron, leading to $\epsilon_U=\eta_U=\rho_U=0$ in the coordinate of $(\boldsymbol{e}_1,\,\boldsymbol{e}_2)$ (see Figure \ref{fig:e1e2e3}).

For a homogeneous, stationary medium, the general solution of the transfer equation for a weakly anisotropic dielectric tensor is
\begin{equation}
\vec{S}=\vec{\varepsilon}s+\vec{X}\exp(-\tau)+\vec{C}_0.
\end{equation}
Different solutions have been obtained for various limiting cases. For instance, a no-absorption solution is given by \cite{1975ApJ...196..125P}, and another solution that converges to a non-zero finite value at $\tau\rightarrow\infty$ was derived by \cite{1977ApJ...214..522J}.
These limiting cases do not apply to FRBs, which are extremely bright single radio pulses.

Conservatively, one estimates that the bright temperature of an FRB typically exceeds $10^{35}$ K at GHz band \citep{2016MPLA...3130013K,2018PhyU...61..965P}.
In a thermal plasma, the absorption coefficients can be calculated according to Kirchhoff's law, i.e., $\eta_S=\epsilon_S/B_\nu$, where $B_\nu$ is the Planck function.
Therefore, the initial incoming radiation has $S_0/B_\nu=\eta_S S_0/\epsilon_S\gg1$ (the superscript ``0'' for $s=0$).
The vector of $\vec{\varepsilon}$ can be neglected, equivalent to a no-emission medium.

The observed Stokes vectors are referenced to the coordinate frame $(\boldsymbol{a},\,\boldsymbol{b})$, where $\boldsymbol{a}$ corresponds to the north at
the observer and $\boldsymbol{b}$ corresponds to the east (see Figure \ref{fig:e1e2e3}). The rotation matrix can be written as
\begin{equation}
\boldsymbol{R}(\chi_p)=\left(\begin{array}{cccc}
1 & 0 & 0 & 0 \\
0 & \cos 2 \chi_p & \sin 2 \chi_p & 0 \\
0 & -\sin 2 \chi_p & \cos 2 \chi_p & 0 \\
0 & 0 & 0 & 1
\end{array}\right).
\end{equation}
Therefore one can obtain
\begin{equation}
\frac{d}{ds} \vec{S}^{\prime}=-\boldsymbol{M}^{\prime} \vec{S}^{\prime},
\end{equation}
\begin{equation}
\begin{aligned}
\vec{S'}&=\boldsymbol{R}(\chi_p)\vec{S}=\left(\begin{array}{l}
I \\
Q\cos2\chi_p+U\sin2\chi_p \\
U\cos2\chi_p-Q\sin2\chi_p \\
V
\end{array}\right),\\
\boldsymbol{M'}&\equiv\left(\begin{array}{cccc}
\eta_I' & \eta_Q' & \eta_U' & \eta_V' \\
\eta_Q' & \eta_I' & \rho_V' & -\rho_U' \\
\eta_U' & -\rho_V' & \eta_I' & \rho_Q' \\
\eta_V' & \rho_U' & -\rho_Q' & \eta_I'
\end{array}\right)
=\boldsymbol{R}(\chi_p)\boldsymbol{M}\boldsymbol{R}(-\chi_p).
\end{aligned}
\end{equation}
With such an absence of emissivity, the solution of the transfer equation can be written as
\begin{widetext}
\[
\begin{aligned}
\vec{S'}&=\exp(-\tau)\\
&\times\left[\left(\begin{array}{cccc}
\frac{1}{2}\left(1+q^2+u^2+v^2\right) & -\boldsymbol{u} \times \boldsymbol{v} &-\boldsymbol{v} \times \boldsymbol{q} & -\boldsymbol{q} \times \boldsymbol{u} \\
\boldsymbol{u} \times \boldsymbol{v} & \frac{1}{2}\left(1+q^2-u^2-v^2\right) & \boldsymbol{q} \cdot \boldsymbol{u} &\boldsymbol{q} \cdot \boldsymbol{v} \\
\boldsymbol{v} \times \boldsymbol{q} & \boldsymbol{q} \cdot \boldsymbol{u}  & \frac{1}{2}\left(1-q^2+u^2-v^2\right) & \boldsymbol{u} \cdot \boldsymbol{v}  \\
\boldsymbol{q} \times \boldsymbol{u} &\boldsymbol{q} \cdot \boldsymbol{v} & \boldsymbol{u} \cdot \boldsymbol{v}  & \frac{1}{2}\left(1-q^2-u^2+v^2\right)
\end{array} \right)\, \cosh (\kappa s)\right. \\
& -\left(\begin{array}{cccc}
0 &\boldsymbol{q} \cdot \boldsymbol{k} & \boldsymbol{u} \cdot \boldsymbol{k}  & \boldsymbol{v} \cdot \boldsymbol{k} \\
\boldsymbol{q} \cdot \boldsymbol{k} & 0 & -\boldsymbol{v} \times \boldsymbol{k} & \boldsymbol{u} \times \boldsymbol{k}  \\
\boldsymbol{u} \cdot \boldsymbol{k}  & \boldsymbol{v} \times \boldsymbol{k} & 0 & -\boldsymbol{q} \times \boldsymbol{k} \\
\boldsymbol{v} \cdot \boldsymbol{k} & -\boldsymbol{u} \times \boldsymbol{k}  &\boldsymbol{q} \times \boldsymbol{k} & 0
\end{array}\right) \sinh (\kappa s) \\
& +\left(\begin{array}{cccc}
\frac{1}{2}\left(1-q^2-u^2-v^2\right) & -\boldsymbol{v} \times \boldsymbol{u} & -\boldsymbol{q} \times \boldsymbol{v} & -\boldsymbol{u} \times \boldsymbol{q} \\
\boldsymbol{v} \times \boldsymbol{u} & \frac{1}{2}\left(1-q^2+u^2+v^2\right) & -\boldsymbol{q} \cdot \boldsymbol{u} & -\boldsymbol{q} \cdot \boldsymbol{v} \\
\boldsymbol{q} \times \boldsymbol{v} & -\boldsymbol{q} \cdot \boldsymbol{u} & \frac{1}{2}\left(1+q^2-u^2+v^2\right) & -\boldsymbol{u} \cdot \boldsymbol{v} \\
\boldsymbol{u} \times \boldsymbol{q} & -\boldsymbol{q} \cdot \boldsymbol{v} & -\boldsymbol{u} \cdot \boldsymbol{v} & \frac{1}{2}\left(1+q^2+u^2-v^2\right)
\end{array}\right) \cos \left(\kappa_* s\right) \\
& \left.-\left(
\begin{array}{cccc}
0 &\boldsymbol{q} \times \boldsymbol{k} & \boldsymbol{u} \times \boldsymbol{k} & \boldsymbol{v} \times \boldsymbol{k} \\
\boldsymbol{q} \times \boldsymbol{k} & 0 & \boldsymbol{v} \cdot \boldsymbol{k} & -\boldsymbol{u} \cdot \boldsymbol{k} \\
\boldsymbol{u} \times \boldsymbol{k} & -\boldsymbol{v} \cdot \boldsymbol{k} & 0 &\boldsymbol{q} \cdot \boldsymbol{k} \\
\boldsymbol{v} \times \boldsymbol{k} & \boldsymbol{u} \cdot \boldsymbol{k} & -\boldsymbol{q} \cdot \boldsymbol{k} & 0
\end{array}\right) \sin \left(\kappa_* s\right)\right]\left(\begin{array}{l}
I_0 \\
Q_0\cos2\chi_p+U_0\sin2\chi_p \\
U_0\cos2\chi_p-Q_0\sin2\chi_p \\
V_0
\end{array}\right).\\
&
\end{aligned}
\label{eq:solution}
\]
\end{widetext}
The symbols $\boldsymbol{\zeta}$, $\boldsymbol{\zeta_*}$, $\boldsymbol{q}$, $\boldsymbol{v}$, and $\boldsymbol{k}$ represent two-vectors defined as
\begin{equation}
\begin{aligned}
\boldsymbol{\zeta} &\equiv\left(\eta_Q' ,\eta_U',  \eta_V'\right), \\
\boldsymbol{\zeta_*} &\equiv\left(\rho_Q', \eta_U' , \rho_V'\right), \\
\boldsymbol{q} &\equiv\left(\eta_Q', \rho_Q'\right)\left[\kappa^2+\kappa_*^2\right]^{-1 / 2}, \\
\boldsymbol{u}& \equiv\left(\eta_U', \rho_U'\right)\left[\kappa^2+\kappa_*^2\right]^{-1 / 2}, \\
\boldsymbol{v}& \equiv\left(\eta_V', \rho_V'\right)\left[\kappa^2+\kappa_*^2\right]^{-1 / 2}, \\
\boldsymbol{k} &\equiv\left(\kappa, \kappa_*\right)\left[\kappa^2+\kappa_*^2\right]^{-1 / 2},
\end{aligned}
\end{equation}
and the two parameters $\kappa$ and $\kappa_*$ denote 
\begin{equation}
\begin{aligned}
&\kappa =\frac{1}{\sqrt2}\left\{\left[\left(\zeta^2-\zeta_*^2\right)^2+4\left(\boldsymbol{\zeta}\cdot \boldsymbol{\zeta_*}\right)^2\right]^{1 / 2}+\zeta^2-\zeta_*^2\right\} ^{1 / 2}, \\
&\kappa_* =\frac{1}{\sqrt2}\left\{\left[\left(\zeta_*^2-\zeta^2\right)^2+4\left(\boldsymbol{\zeta}\cdot \boldsymbol{\zeta_*}\right)^2\right]^{1 / 2}+\zeta_*^2-\zeta^2\right\}^{1 / 2}.
\end{aligned}
\end{equation}

\subsection{Thermal Plasma Medium}\label{sec2.2}

\begin{figure}
\plotone{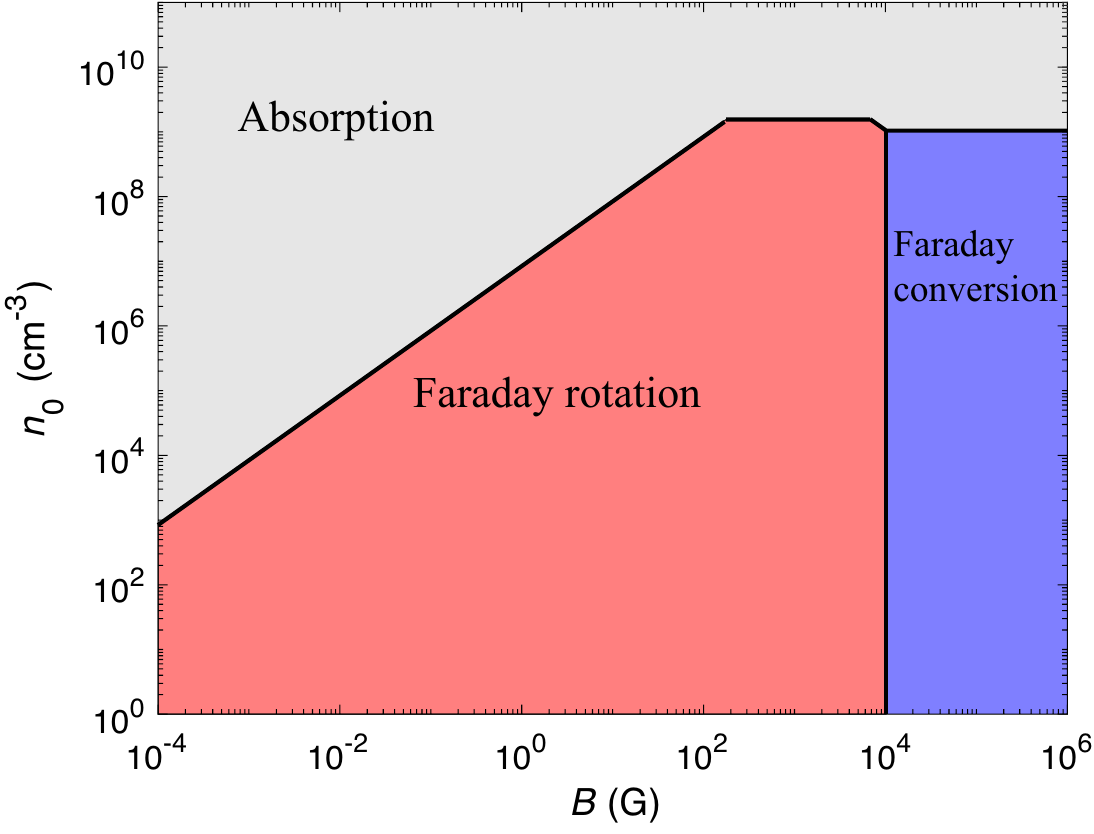}
\plotone{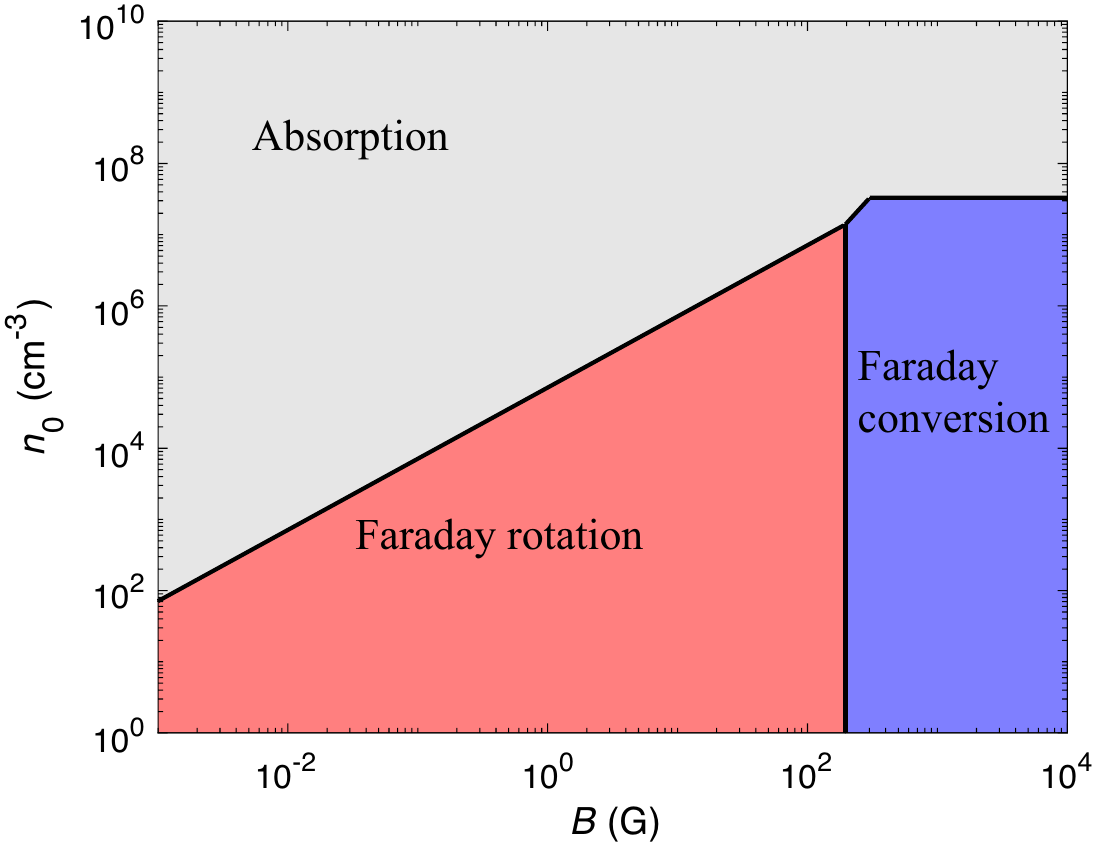}
\caption{Dominant regions of the transformation coefficients for a thermally distributed plasma: Top plane for $\theta_B=15^\circ$; Bottom plane for $\theta_B=75^\circ$.
\label{fig:thermal}}
\end{figure}

\begin{figure}
\plotone{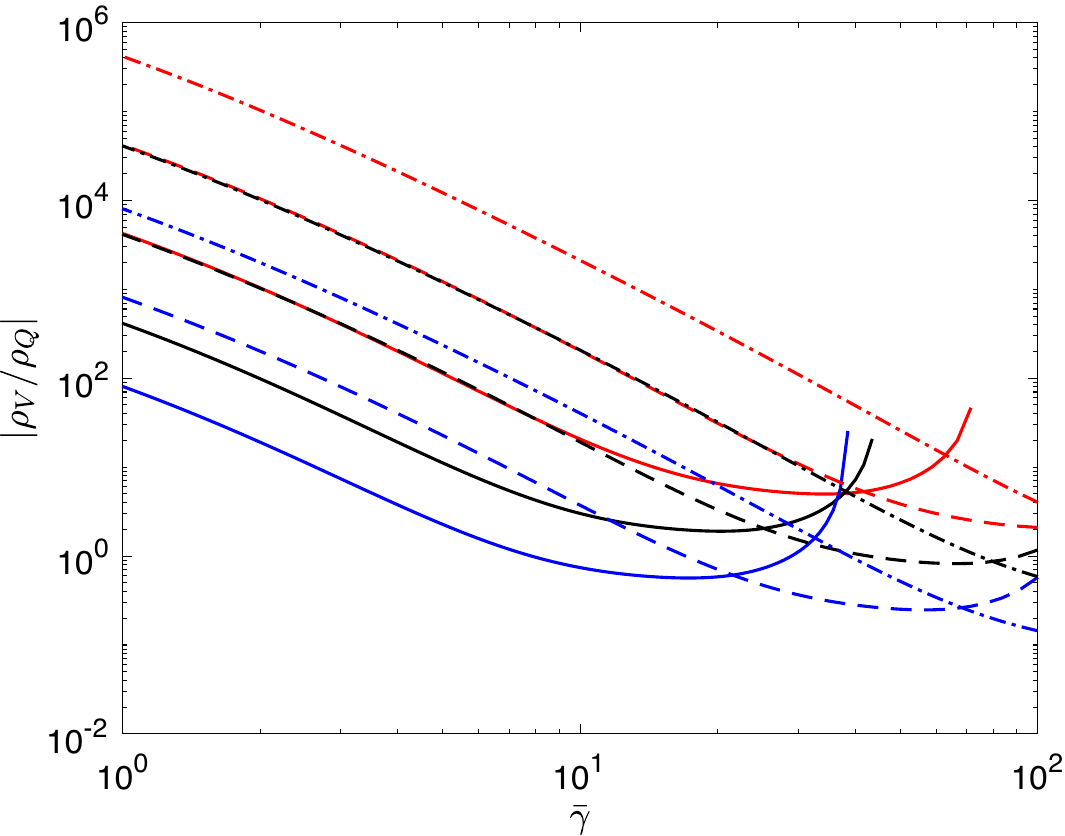}
\caption{Relationship between $|\rho_V/\rho_Q|$ and $\bar\gamma$ for a relativistic plasma. The value of $|\rho_V/\rho_Q|$ reflects whether Faraday rotation or conversion dominates the propagation. Different colors denote different $\theta_B$ values: red ($\theta_B=15^\circ$), black ($\theta_B=45^\circ$), blue ($\theta_B=75^\circ$). The magnetic field is taken as $B=10^{-3}$ G (dashed-dotted lines), $B=10^{-2}$ G (dashed lines), and $B=10^{-1}$ G (solid lines), respectively.
\label{fig:rpara}}
\end{figure}

Collisions between electrons and ions in a plasma can generate Bremsstrahlung.
In a magnetized plasma, gyration of relativistic charged particles can power synchrotron radiation. For a single charged particle, the ratio of power between synchrotron and bremsstrahlung can be estimated as \citep{2018pgrb.book.....Z}
\begin{equation}
\frac{P_{\rm syn}} {P_{\rm ff} }\sim \frac{\gamma m_p}{\alpha_em_e}\beta^2  \sigma_m \sim 2.5 \times 10^5 \beta^2\gamma \sigma_m,
\end{equation}
where $\gamma$ is the electron Lorentz factor, $\beta$ is the velocity normalized to the speed of light $c$, $m_p$ is the mass of proton, $m_e$ is the mass of electron, $\alpha_e\approx1/137$ is the fine-structure constant, and $\sigma_m = B^2 / (8\pi\rho c^2)$ is the magnetization parameter, $B$ is the magnetic field strength, and $\rho$ is the mass density. 
The synchrotron radiation and absorption are significant as long as $\sigma_m$ is not too small. One can see bremsstrahlung and free-free absorption would dominate only when the plasma is very dense ($\rho c^2 \gg B^2/8\pi$) and non-relativistic ($\beta\ll1$).

We consider a thermally distributed plasma, whose number density is
\begin{equation}
n(E)=n_0\frac{E^2}{2(k_BT)^3}\exp(-E/k_BT),
\end{equation}
where $E$ is energy, $k_B$ is the Boltzmann constant and $T$ is the temperature, respectively.
GHz radio waves have $h\nu\ll k_BT$ so that $B_\nu$ works in the Rayleigh-Jeans regime.
The ensuing discussion will primarily address the scenario under $\rho c^2 \gg B^2/8\pi$, which may be more applicable to the interstellar medium with weak magnetic fields or stellar wind environments.
In such cases, the free-free absorption, i.e., the inverse process of Bremsstrahlung, would become the dominant absorption mechanism.

In a cold plasma, the absorption coefficient of $I$ can be calculated due to the Kirchhoff's law \citep{1965ARA&A...3..297G,1969SvA....13..396S,1979rpa..book.....R}:
\begin{equation}
\eta_I=\frac{8e^6n_0^2}{3\sqrt{2\pi}(k_BTm_e)^{3/2}c\nu^2}\ln{\left[\frac{(2k_BT)^{3/2}}{4.2\pi e^2m_e^{1/2}\nu}\right]},
\label{eq:etaI}
\end{equation}
where $e$ is elementary charge unit and $\nu=\omega/(2\pi)$ is frequency.
The absorption coefficients for linear and circular polarizations exhibit slight differences, because of the distinct interaction cross-sections of radiation polarized along and across the magnetic field. The magnetic field in interstellar space can hardly exceed $\sim100$ G except for stellar atmospheres or stellar winds from compact stars.
The gyrofrequency $\nu_B=eB/(2\pi m_ec)=2.8\times10^6(B/1\,\rm G)$ is therefore usually much smaller than the wave frequency.
For $\nu_B\ll\nu$, the absorption coefficients can be given by \citep{1969SvA....13..396S}
\begin{equation}
\begin{aligned}
&\eta_Q=\frac{3}{2}\left(\frac{\nu_B\sin\theta_B}{\nu}\right)^2\eta_I,\\
&\eta_V=-\frac{2\nu_B\cos\theta_B}{\nu}\eta_I,
\end{aligned}
\label{eq:etaQV}
\end{equation}
where $\theta_B$ is the angle between the magnetic field $\vec{B}$ and wave vector.
For $\nu\ll\nu_B$, the opacity of radiation polarized along the magnetic field is multiplied by a factor of 3/2, while the opacity for radiation polarized perpendicular to the magnetic field is enhanced by a factor of $\frac{1}{2}\ln[\nu_B/(4.85\nu)]$ \citep{1975ApJ...196..125P}.
This enhancement of the opacity of perpendicular photons produce a slightly increase in the absorption of the circularly polarized waves.

Faraday rotation and conversion coefficients can be read off from the plasma response tensor.
A generalized form of Faraday rotation and conversion is given by \citep{2008ApJ...688..695S,2011MNRAS.416.2574H}
\begin{equation}
\begin{aligned}
&\rho_V=\frac{2e^2\nu_Bn_0\cos\theta_B}{\nu^2m_ec}\left[\frac{K_0(\bar\gamma^{-1})}{K_2(\bar\gamma^{-1})}\right]g(X),\\
&\rho_Q=-\frac{n_0e^2\nu_B^2\sin^2\theta_B}{\nu^3m_ec}\left[\frac{K_1(\bar\gamma^{-1})}{K_2(\bar\gamma^{-1})}+6\bar\gamma\right]f(X),
\end{aligned}
\label{eq:rho_thermal}
\end{equation}
where $\bar\gamma=k_BT/(m_ec^2)$ represents the characteristic Lorentz factor, and the two multipliers $f(X)$ and $g(X)$ are defined in Appendix \ref{response}.
The parameter $X$ is much smaller than unity even for $\nu\approx\nu_B$ when $\bar\gamma\ll1$, i.e., a cold plasma case, leading to $f(X)\approx g(X)\approx1$.

For most astrophysical environments, the propagation of electromagnetic waves is primarily governed by absorption or one of the Faraday effects.
A characteristic absorption coefficient is defined as $\eta=\max[\eta_I,\,\eta_Q,\,\eta_V]$.
We investigate the dominant regions for the propagation coefficients, and define the corresponding effect as dominant when either the characteristic absorption or mixing Faraday coefficients exceed all others.
We compare $\eta$ and the mixing Faraday coefficients as functions of number density and magnetic field strength with $T=1$ K and $\nu=10^9$ Hz.
The dominant regions for $\theta_B=15^\circ$ and $\theta_B=75^\circ$, which are essentially determined by $\max[\eta,\,\rho_V,\, \rho_Q]$, are plotted in Figure \ref{fig:thermal}.
The absorption for such conditions is mainly dominated by the free-free absorption.
For a low-density medium, the propagation effects are dominated by Faraday rotation if $B\lesssim3\times10^2(\cos\theta_B/\sin^2\theta_B)$ G, while a Faraday conversion becomes strong for $B\gtrsim3\times10^2(\cos\theta_B/\sin^2\theta_B)$ G.
When the medium has a sufficiently high density density, the absorption coefficient becomes polarization-dependent as the magnetic field strength increases.
This leads to the propagation coefficient, $\eta$, being governed by the relative magnitudes of the absorption coefficients for linearly or circularly polarized radiation rather than $\eta_I$.

For a hot plasma, we mainly focus on the scenarios of $\nu_B\ll\nu$, which is the typical condition in the ISM. Also, there are complex higher order effects when $\nu_B\gg\nu$.
Even with such a weak magnetic field regime, synchrotron absorption can dominate unless the plasma is extremely dense.
The elliptical polarization property of synchrotron radiation enables different interactions between electrons and photons, e.g., absorption, with different polarization modes. A magnetized plasma medium can also lead to mixing conversion of Stokes basis, i.e., Faraday rotation and conversion.

In a hot thermal plasma, Faraday mixing coefficients are larger than $\eta$ for $\nu_B\ll\nu$.
Therefore, we investigate $|\rho_V/\rho_Q|$ as a function of $\bar\gamma$ with different conditions, as shown in Figure \ref{fig:rpara}, rather than the dominant regions as Section \ref{sec2.2}.
The Faraday conversion can be more significant than Faraday rotation when $10\lesssim\bar\gamma$ and $B\lesssim10^{-2}$ G.
A more relativistic plasma tends to have a larger Faraday conversion coefficient due to the second term in Equation (\ref{eq:rho_thermal}).
The relativistic effect makes Faraday rotation of a hot plasma smaller compared to a cold plasma.

\section{Spectro-polarimetric properties}\label{sec3}
A magnetized plasma medium responds differently to waves with different frequencies so that the outgoing waves have noticeable differences in spectro-polarimetric properties with incoming waves.
We assume that the intrinsic FRB emissions are highly polarized, and their Stokes parameters are independent of frequency.
In the following discussions, we mainly focus on the thermal plasma case.
We consider the medium to have a strong magnetic field or large number density so that the waves may have complex spectro-polarimetric properties.
The medium with significant Faraday rotation and comparable conversion is considered in Section \ref{sec3.1}, and that for Faraday rotation and absorption is investigated in Section \ref{sec3.2}.

\subsection{Faraday mixing spectra}\label{sec3.1}

\begin{figure*}[ht!]
\plotone{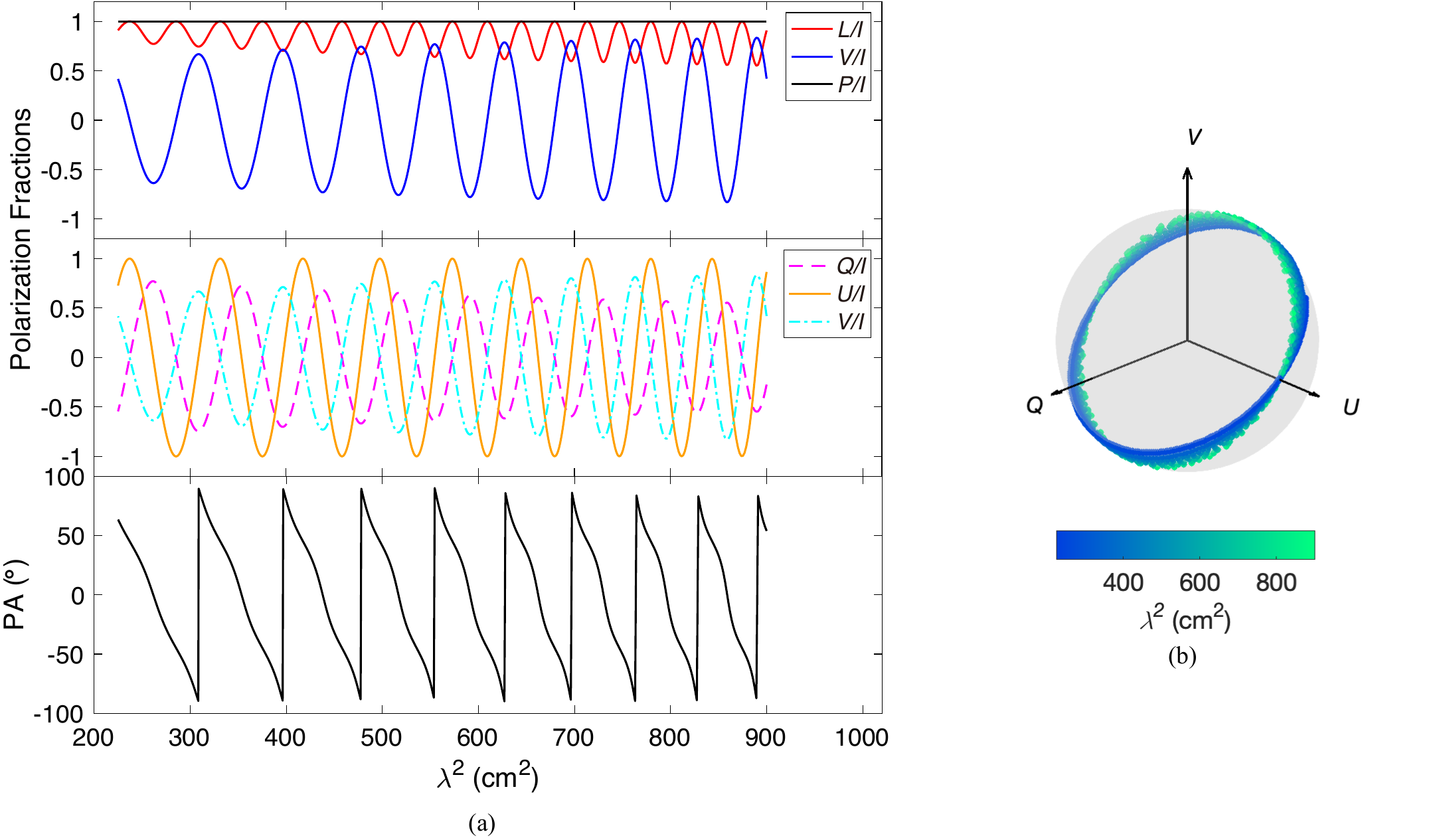}
\caption{Simulated spectro-polarization profiles. (a) Top panel: linear, circular and total polarization fractions; Middle panel: $Q/I$, $U/I$, and $V/I$. Bottom panel: polarization angle. (b) Poincar\'e sphere representation of the spectropolametric properties. The parameters are adopted as $\theta_B=105^\circ$, $\chi_p=0$, $L=10^{13}$ cm, $B=3\times10^2$ G, $n_0=1\,\rm cm^{-3}$, $T=1$ K, $Q_0/I_0=0$, $U_0/I_0=1$, $V_0/I_0=0$. 
\label{fig:FRFC1}}
\end{figure*}

\begin{figure*}[ht!]
\plotone{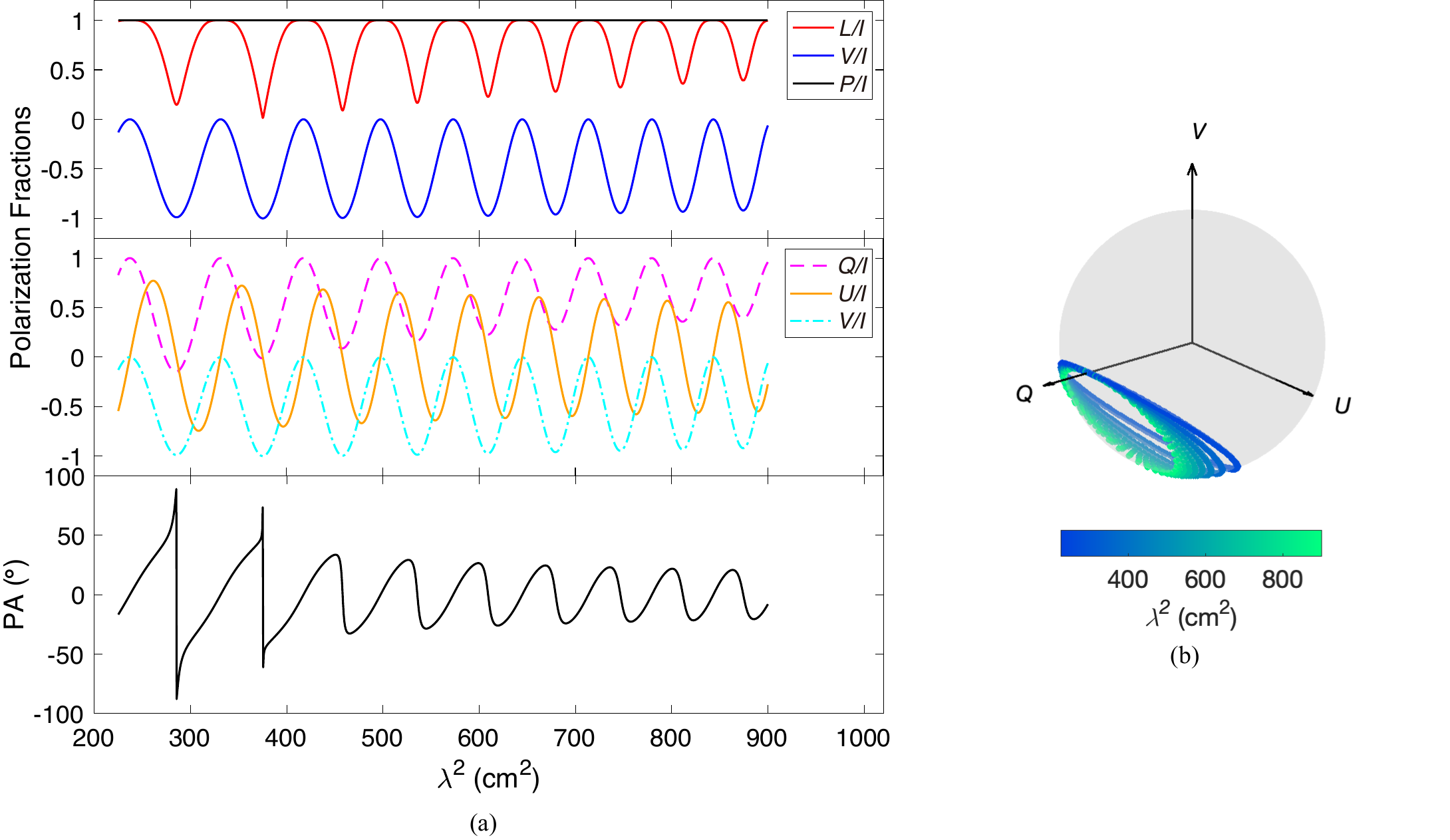}
\caption{Same as Figure \ref{fig:FRFC1}, but for $\theta_B=75^\circ$, $Q_0/I_0=1$, $U_0/I_0=0$, $V_0/I_0=0$. 
\label{fig:FRFC2}}
\end{figure*}

If the magnetic field is parallel to the line of sight (LOS), the intrinsic plasma modes are circularly polarized but with different refractive indices.
Birefringence of those two modes causes the plane of any linear polarization to rotate, that is, the pure Faraday rotation.
The change of PA has a determined frequency dependence, i.e., $\rm{PA=RM}\lambda^2$, where $\lambda$ is the wavelength.
In this case, the Stokes spectra rotate around the $V$-axis in a Poincar\'e sphere, and the linear and circular polarizations do not change.
However, if $B$-field is strictly perpendicular to the LOS, the intrinsic plasma modes are linearly polarized. Different propagation speeds of the two eigen modes (O-mode and X-mode) lead to a partial conversion of linear polarization into circular polarization.
The spectra then rotate around an axis in the $Q-U$ plane.
In reality, the $B$-field is likely not strictly perpendicular to the LOS, and the presence of the parallel component of the $B$-field enables a Faraday rotation.
Consequently, the medium with significant Faraday conversion is always associated with Faraday rotation.

We consider a cold plasma with comparable Faraday rotation and conversion coefficients.
The introduction of the $\chi_p$ is due to the selection of the instrument coordinates and is not determined by the intrinsic physics of the plasma medium.
Therefore, we take $\chi_p=0$ in the following simulations.
According to Figure \ref{fig:thermal}, when $B\sim3\times10^2$ G, indicating $\nu\sim\nu_B$, the condition of $|\rho_V|\approx|\rho_Q|$ is satisfied.
Assume that the plasma number density is equivalent to that of the interstellar medium, i.e. $n_0=1\,\rm cm^{-3}$ \citep{1993PASP..105.1127G}, and the scale is $\sim1$ AU. 
The cold plasma requires $k_BT \ll m_ec^2$, and at this point, the Faraday mixing coefficients are independent of temperature (see Equation (\ref{eq:rho_cold})), hence we set $T=1$ K for simplicity.
We take $L=10^{13}$ cm, $B=3\times10^2$ G, and $\theta_B=105^\circ$, indicating that the projection of the $B$-field on the LOS has an opposite direction with respect to the LOS.
For these parameters, the plasma medium has significant rotation and conversion rates and a negative value of RM.
The medium provides a dispersion measure (DM) of $10^{-5}\,\rm{cm^{-2}}$ and RM of $-340.6\,\rm{rad\,m^{-2}}$.

We simulate the polarization spectrum by assuming an incoming wave with $Q_0/I_0=0$, $U_0/I_0=1$, and $V_0/I_0=0$ for interest, as shown in Figure \ref{fig:FRFC1}.
The linear polarization component is $L=\sqrt{Q^2+U^2}$ with a corresponding PA defined as $1/2\arctan(U/Q)$.
The total polarization component is $P=\sqrt{Q^2+U^2+V^2}$.
All of $Q$, $U$, and $V$ oscillate with frequency.
$Q$ has a phase that is $\pi/2$ earlier than $U$ but has an opposite phase with $V$.
The oscillation frequency of linear polarization is approximately twice of that of circular polarization.
The spectra no longer rotate around the $V$-axis or an axis within the $Q-U$ plane on the Poincar\'e sphere.
Because of the different frequency dependencies between Faraday rotation and conversion, the rotational axis of the polarization spectrum undergoes precession as the frequency changes.
The PA spectrum within one period can be characterized as an `S-shape' rather than a linear function of $\rm{PA=RM}\lambda^2$.
The polarization spectrum resembles a GFR spectrum with spectral index $\alpha>2$, which is intrinsically the Faraday mixing case.

\begin{figure*}
\plotone{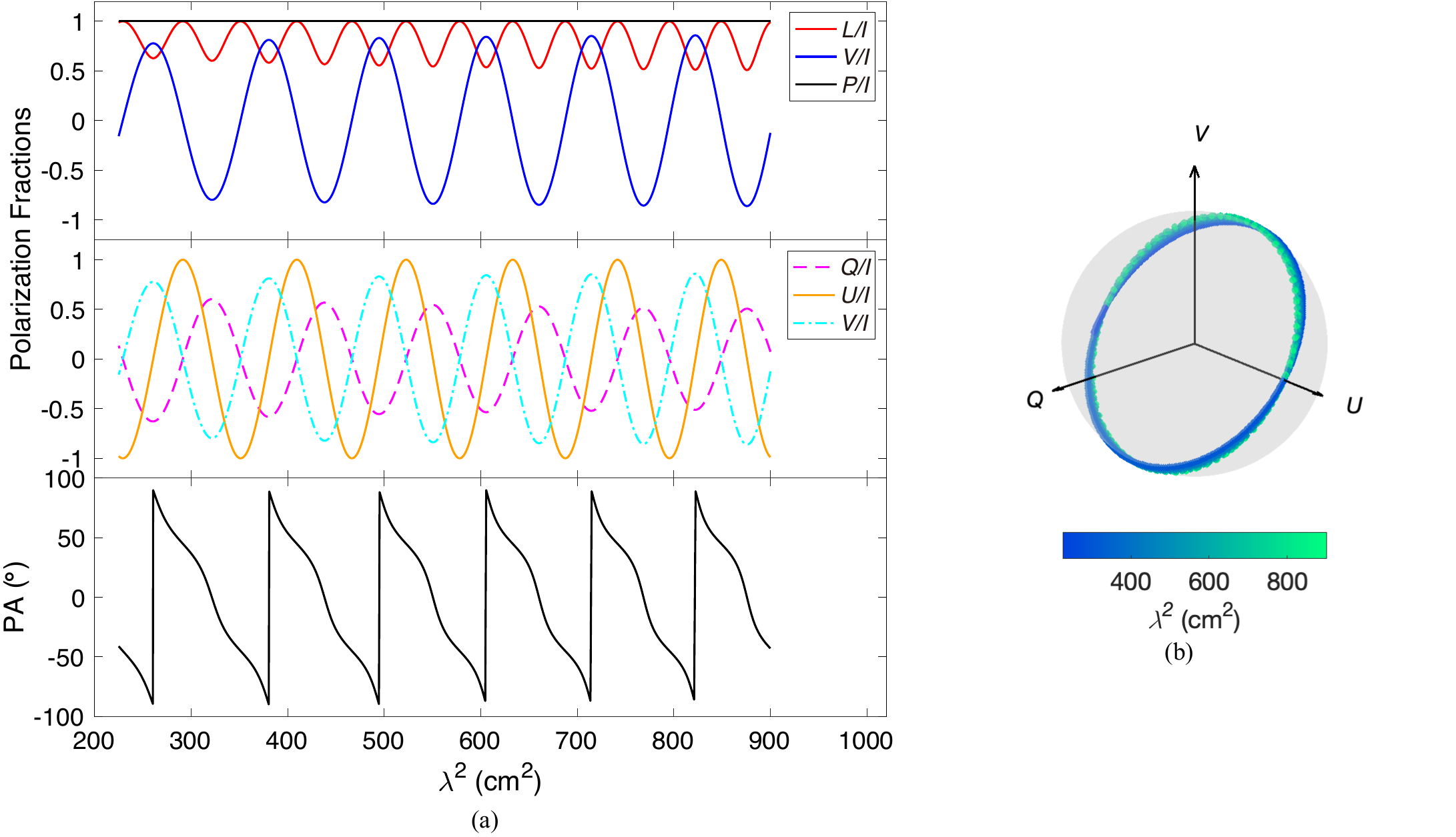}
\caption{Same as Figure \ref{fig:FRFC1}, but for a relativistic case with $\bar\gamma=10^2$, $\theta_B=135^\circ$ $B=10^{-3}$ G, $n_0=3\times10^3\,\rm cm^{-3}$, and $L=10^{18}$ cm.
\label{fig:FRFC3}}
\end{figure*}

Consider a different incoming wave with a different initial PA.
We change the incoming wave to $Q_0/I_0=1$, $U_0/I_0=0$, $V_0/I_0=0$, and the angle as $\theta_B=75^\circ$.
The absolute value of RM is the same as Figure \ref{fig:FRFC1} but for an opposite sign.
The simulation of the outgoing polarization spectra is shown in Figure \ref{fig:FRFC2}.
Same as Figure \ref{fig:FRFC1}, the polarization spectra have precession on the Poincar\'e sphere.
However, in the $\theta_B<\pi/2$ regime, the phase of $Q$ is later than that of $U$ by a factor of $\pi/2$, and $V$ shares the same phase as $Q$.
A similar precession could happen.
Compared to Figure \ref{fig:FRFC3}, the direction of the precession on the Poincar\'e sphere is reversed due to an opposite Faraday rotation.
Different incoming wave Stokes determine the quadrants in the Poincar\'e sphere.
The oscillations in U are confined within the range of $\pm|U_0|$.
The oscillation amplitude of $Q$ decreases with decreasing frequency, and its average value is maintained at $Q_0$.
If the $Q_0/I_0$ is close to 1 and $U_0/I_0$ is small, the polarization spectra are hard to cross the $U$-axis so the oscillation amplitude of PA is smaller than $90^\circ$.

The intrinsic plasma modes for ultra-relativistic plasma are linearly polarized.
The propagation-induced phase delay of the modes produces a partial conversion of linear into circular polarization.
Therefore, compared to the cold plasma, the Faraday rotation rate of the hot plasma decreases while the Faraday conversion rate increases.
According to Figure \ref{fig:rpara}, we chose a set of parameters $\bar\gamma=10^2$, $\theta_B=135^\circ$ and $B=10^{-3}$ G that meet the condition of $|\rho_V|\approx|\rho_Q|$.
The polarization spectrum with the same incoming wave as Figure \ref{fig:FRFC1} is plotted in Figure \ref{fig:FRFC3}.
The hot plasma shares similar polarization properties with the cold plasma in Figure \ref{fig:FRFC1}.

A hot plasma requires a larger $n_0L$ to maintain a sufficient RM, allowing the spectrum to oscillate frequently compared with the cold plasma case.
The DM of a relativistic plasma is modified as
\begin{equation}
\mathrm{DM}=\int n_0 K_1(\bar\gamma^{-1})/K_2(\bar\gamma^{-1})ds.
\end{equation}
For $\bar\gamma\gg1$, the DM is approximated to $\int n_0\bar\gamma^{-1}ds/2$.
The scenario of Figure \ref{fig:FRFC3} gives a DM of $\sim5\,\rm{pc\,cm^{-3}}$.

Faraday conversion could be significant even for a small $B$-field and $\theta_B$ in a hot plasma.
Based on Equation (\ref{eq:rho_thermal}), we find that the hot plasma can share similar spectra with the cold plasma for $X\lesssim1$ when the following condition is satisfied:
\begin{equation}
\frac{B_c\sin^2\theta_{B,c}}{\cos\theta_{B,c}}\simeq\frac{12\bar\gamma^{3}}{\ln\bar\gamma}\frac{B_h\sin^2\theta_{B,h}}{\cos\theta_{B,h}},
\end{equation}
where the subscript `c' denotes the cold plasma and `h' for the hot plasma.
The similarity may make it challenging to distinguish whether the medium is a cold or hot plasma solely based on the polarization energy spectrum.

\subsection{Absorbed Faraday spectra}\label{sec3.2}

\begin{figure*}[ht!]
\plotone{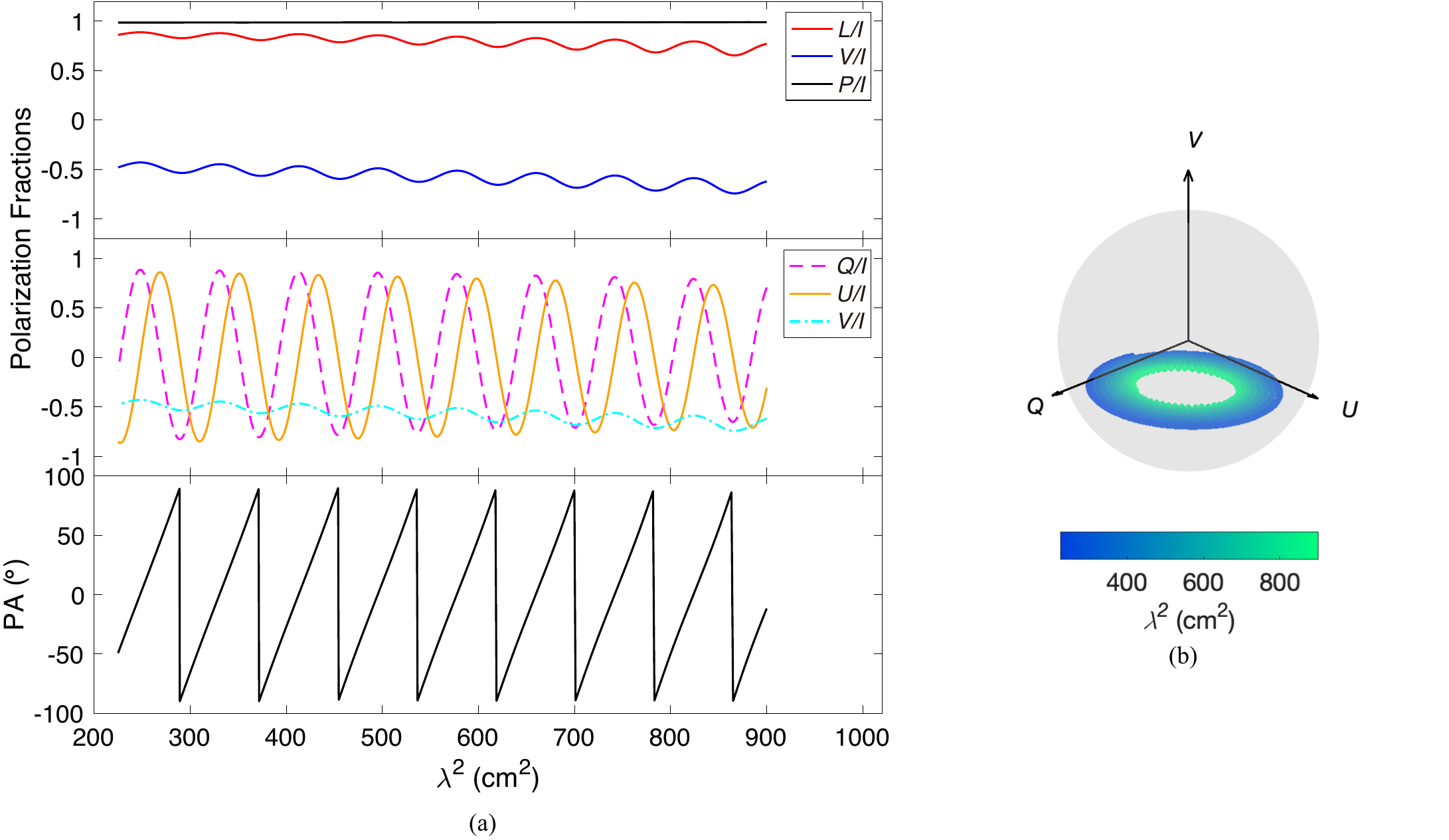}
\caption{Same as Figure \ref{fig:FRFC1} but for $\theta_B=135^\circ$, $L=2\times10^{4}$ cm, $B=10^2$ G, $n_0=10^9\,\rm cm^{-3}$, $T=10^2$ K, $Q_0/I_0=0.9$, $U_0/I_0=0$ and $V_0/I_0=-0.4$.
\label{fig:AbFR2}}
\end{figure*}

The polarization spectra are located on the surface of a Poincar\'e sphere when the emission is 100\% polarized.
If the linear or circular polarization component is absorbed significantly, the polarization spectra would be inside the Poincar\'e sphere.
We consider that the absorption works at a high number density for a cold thermal plasma medium.
The spectra can have both significant absorption and Faraday rotation for a dense magnetized plasma.

Consider a case with significant absorption, Faraday rotation, and conversion.
We set $B=10^2$ G to ensure that the plasma medium possesses comparable Faraday mixing coefficients.
In order to match the absorption coefficient with the Faraday coefficient, the medium requires a high number density and temperature. Thus, we take $n_0=10^9\,\rm cm^{-3}$ and $T=10^2$ K and $L=2\times10^{4}$ cm (e.g., \citealt{1990ApJ...363..554M,2010A&A...518L...1P}), which results in an RM that is roughly consistent with the RMs presented in Figures \ref{fig:FRFC1} and \ref{fig:FRFC2}.
We take $Q_0/I_0=0.9$, $U_0/I_0=0$ and $V_0/I_0=-0.4$ and plot the spectra as shown in Figure \ref{fig:AbFR2}.
The spectrum motion in the Poincar\'e sphere is complex, including precession and radius shrinkage.
Low-frequency waves tend to display a higher degree of circular polarization because linear polarization absorption is higher than circular polarization and is accompanied by some Faraday conversion.
The occurrence of highly circularly polarized bursts possibly leads to an intrinsic mechanism of circular polarization. 

\section{Polarization properties of some FRB sources}\label{sec4}
\subsection{FRB 20180301A}\label{sec4.1}

\begin{figure}[ht!]
\plotone{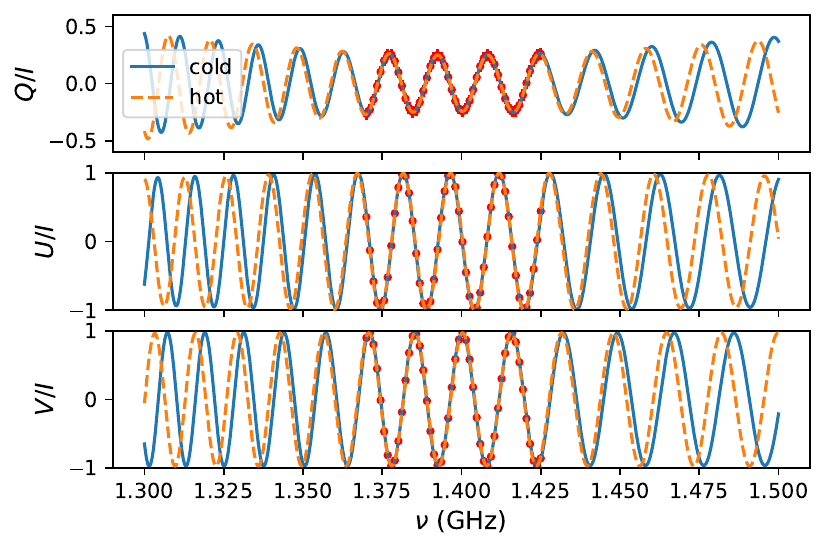}
\caption{The best fitting of cold and hot plasma. The red dots denote the data produced from the best fitting of the FR-GFR model for FRB 20180301A. The blue lines are the Stokes spectra of the cold plasma. The orange lines are the Stokes spectra of the hot plasma.}
\label{fig:GFR}
\end{figure}

\begin{figure*}[ht!]
\plotone{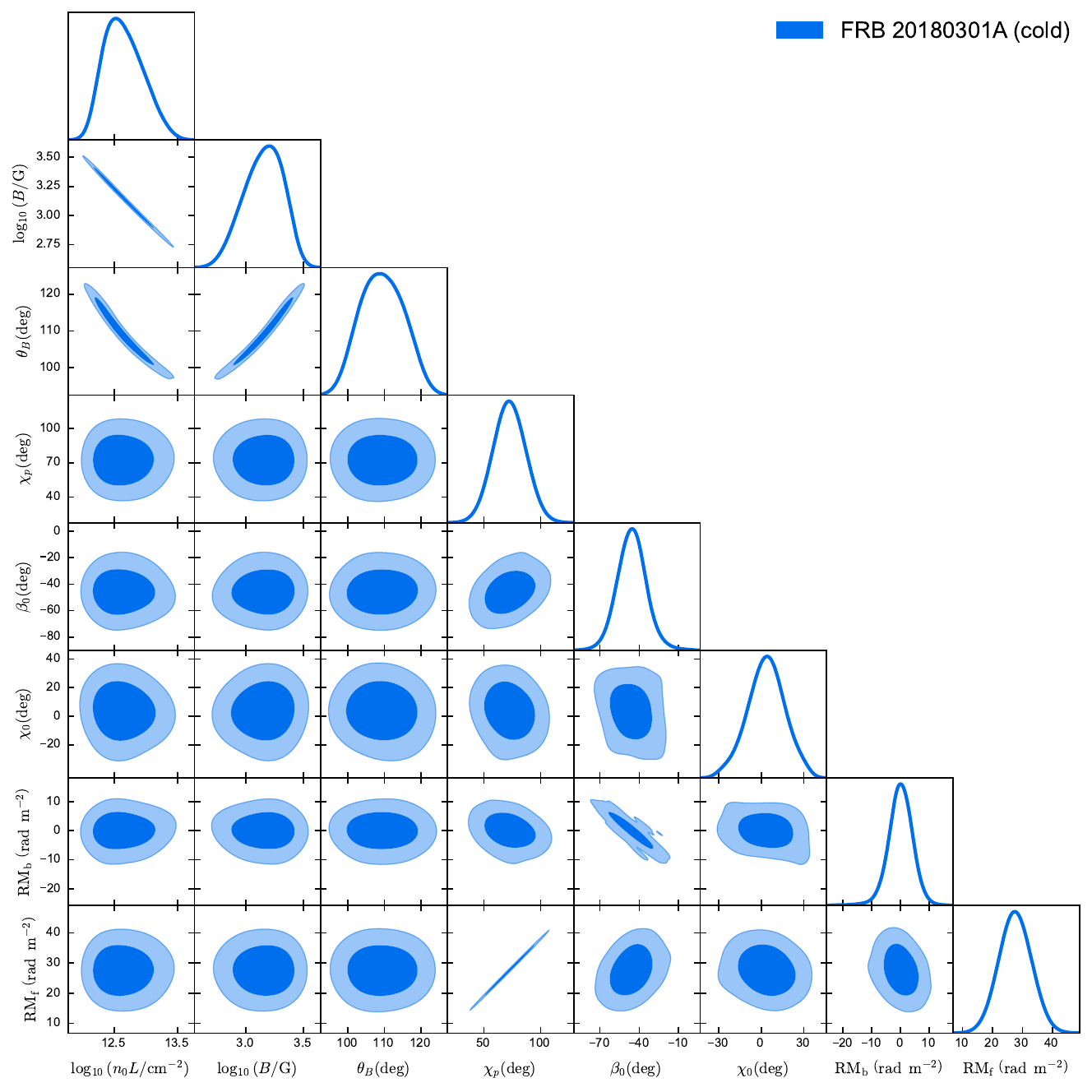}
\caption{The 1D and 2D marginalized probability distributions at 68.3\% and 95.4\% confidence levels for the cold plasma scenario for FRB 20180301A.}
\label{fig:cold}
\end{figure*}

\begin{figure*}[ht!]
\plotone{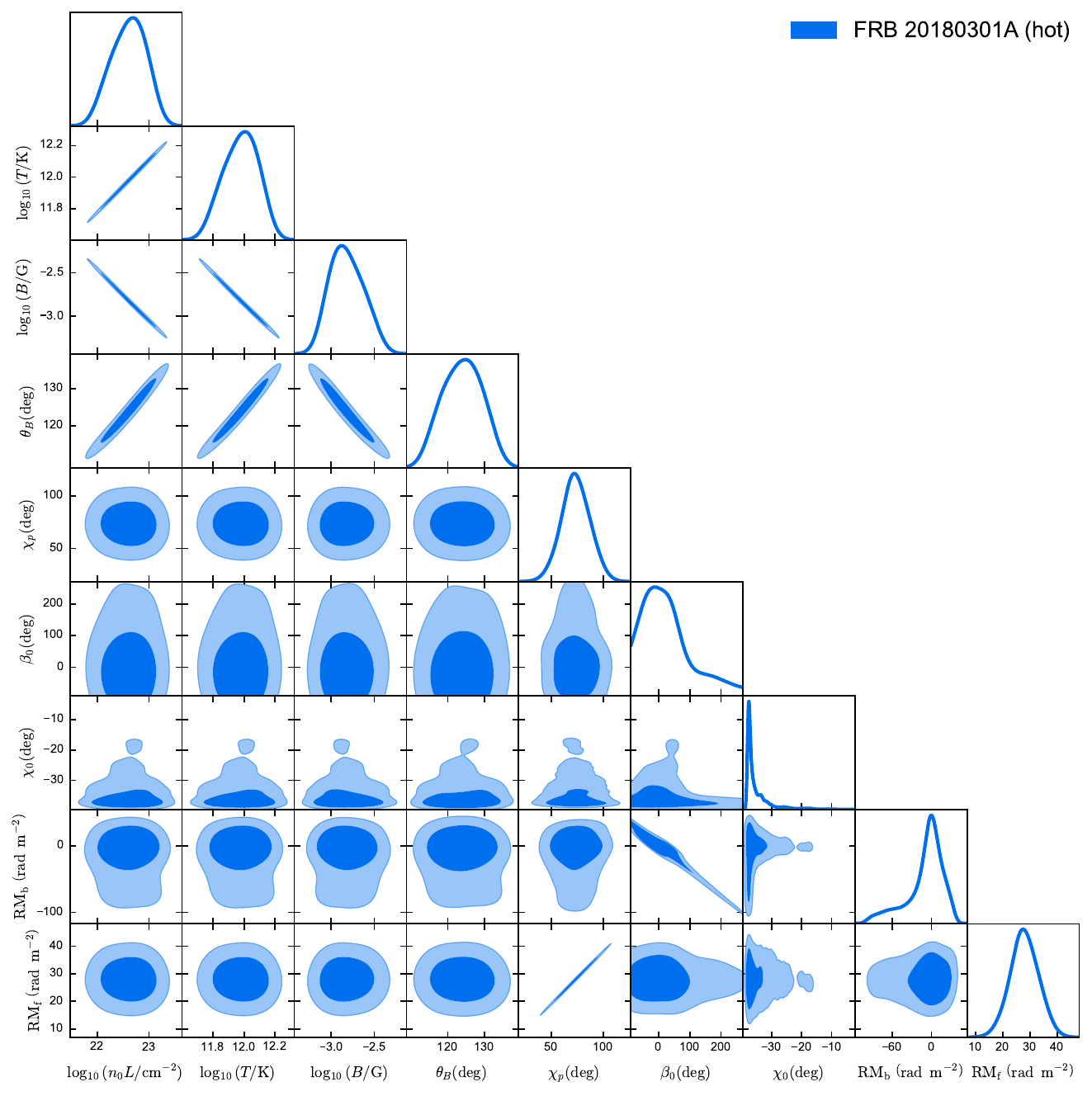}
\caption{Same as Figure \ref{fig:cold} but for the hot plasma scenario.}
\label{fig:hot}
\end{figure*}

FRB 20180301A is the first repeater reported to exhibit variations in PA \citep{2020Natur.586..693L}.
The PA variations are attributed to magnetospheric emission, while the RM variations indicate a complex magnetic environment surrounding the source.
There is one burst of FRB 20180301A that displays frequency-dependent circular polarization along with PA variation behavior \citep{2024MNRAS.534.2485U}.
This Stokes behavior could be modeled using a GFR model.

We argue that our Faraday mixing scenario can effectively represent the GFR model but give a more physical interpretation. The requirement is that one should have comparable coefficients for Faraday rotation and conversion.
We generate a mock GFR spectra of FRB 20180301A using the best-fit parameters in \cite{2024MNRAS.534.2485U}.
The GFR method to retrieve polarization spectra is given in Appendix \ref{apendix:GFR}.
The foreground RM of the GFR model is $27.7\rm\,rad\,m^{-2}$.
The intensity $I$ is set as 1 and the errors ($\sigma$) of $Q$, $U$, and $V$ are set as 0.05.
Considering the identical Gaussian noise in $Q$, $U$, and $V$, one can express the likelihood as
\begin{equation}
    \mathcal{L} \propto \exp\left(-\frac{1}{2} \sum^{N}_{i} \frac{\left( P_i - P_{\mathrm{m},i} \right)^2}{\sigma^2} \right)
\end{equation}
where $P_i$ is the simulated GFR spectra and $P_{\mathrm{m},i}$ is that of the Faraday mixing scenario.

The distance between the source and the medium responsible for the Faraday mixing spectra remains uncertain.
Thus, in addition to considering the presence of a foreground RM contribution layer, we also need to introduce a potential background RM layer.
If the plasma medium is located very close to the FRB source, the RM contribution from this background layer may be negligible.

Since the intensity of the Stokes parameters remains almost unchanged, we neglect all absorption coefficients in our fitting process, which significantly simplifies the fitting procedure.
We found that the total polarization degree of the data remains consistently above 99\%, which validates the exclusion of the absorption effects in our fitting model.
Thus, the total polarization degree is frozen to $P=1$.
We fix the parameter $T$ to 1 K, given that the Faraday mixing coefficients exhibit a negligible temperature dependence and remain effectively constant when $T \lesssim 10^7$ K.

We perform a Bayesian inference using \texttt{emcee} \citep{2013ascl.soft03002F} with uniform priors to sample from the posterior distributions.
We convert the initial Stokes parameters $Q_0$, $U_0$, and $V_0$ to $P$, $\beta_0$, and $\chi_0$.
The relationship between these parameters is referred to as Equation (\ref{eq:Pvector}), in which $\Psi$ is replaced by $\beta_0$ and $\chi$ is replaced by $\chi_0$.
Table \ref{tab1} presents a summary of the constraint results, while Figure \ref{fig:cold} and Figure \ref{fig:hot} illustrate the marginalized posterior densities obtained using \texttt{GetDist} Python package \citep{Lewis:2019xzd} for both hot plasma and cold plasma.
We present the best-fit curves for both cold and hot plasma scenarios, as shown in Figure \ref{fig:GFR}\footnote{The software used for the mixing Faraday analysis is available on https://github.com/GalaxyL777/MixF.}.
The foreground RMs for the cold and hot plasma mediums are ${\rm RM}_f^{\rm cold}=27.5^{+5.5}_{-5.4}\,\rm{rad\,m^{-2}}$ and ${\rm RM}_f^{\rm hot}=27.8^{+5.4}_{-5.0}\,\rm{rad\,m^{-2}}$, and the background RMs for them are ${\rm RM}_b^{\rm cold}=0.9\times10^{-2}\pm4.0\,\rm{rad\,m^{-2}}$ and ${\rm RM}_b^{\rm hot}=-3.1^{+18.1}_{-33.6}\,\rm{rad\,m^{-2}}$.

\begin{table*}
\centering
\caption{68.3\% credible intervals for the parameters of the cold and hot plasma model for FRB 20180301A.}
\begin{tabular}{cccccccc}
\hline\hline
 & $\log_{10}(B/\rm\,G)$ & $\theta_B$ (deg) & $\log_{10}(n_0L/\rm\,cm^{-2})$ & $\log_{10}(T/\rm \,K)$ & $\beta_0$ (deg) & $\chi_0$ (deg) & $\chi_p$ \\
Cold & $3.17^{+0.17}_{-0.19}$ & $109.2^{+6.8}_{-6.4}$ & $12.62^{+0.36}_{-0.29}$ & 0 & $-45.7^{+10.6}_{-10.8}$ & $3.7^{+13.0}_{-13.5}$ & $72.4^{+14.6}_{-14.2}$\\
Hot & $-2.84^{+0.24}_{-0.19}$ & $124.2^{+5.9}_{-6.5}$ & $22.64^{+0.34}_{-0.41}$ & $12.0\pm0.1$ & $14.0^{+85.5}_{-61.0}$ & $-36.8^{+4.6}_{-1.0}$ & $73.0^{+14.3}_{-13.3}$\\
\hline 
 & GRM ($\rm rad\,m^{-\alpha}$) & $\alpha$ & $\lambda_0 ($\rm m$)$ & $\psi$ (deg) & $\chi$ (deg) & $\phi$ (deg) & $\theta$ (deg) \\
GFR & 4351.7 & 2.3 & $0.22$ & $-87.3$ & $-0.1$ & $76.3$ & $104.2$\\
\hline\hline
\end{tabular}
\label{tab1}
\end{table*}

Whether for cold or hot plasmas, the polarization spectra in a specific frequency range can be equivalent to GFR.
In contrast to GFR, which only characterizes the parameters of modulation and rotation on the Poincar\'e sphere, the mixing Faraday scenario reflects the physical characteristics of the medium.
The background RM contribution is negligible in both cold and hot plasma scenarios, suggesting that the plasma medium is likely near the FRB 20180301A source.
The modeled $\theta_B\approx110^\circ$ suggests a $B$-field almost perpendicular to the LOS.
The modeled DM of the cold plasma is $1.4\times10^{-6}\,\rm pc\,cm^{-3}$, and that of the hot plasma is $41.5\,\rm pc\,cm^{-3}$.
The observational DM and RM of FRB 20180301A are $516.76\,\rm{pc\,cm^{-3}}$ and $\sim550\,\rm{rad\,m^{-2}}$ \citep{2020Natur.586..693L}, higher than the modeled DM of both the cold and hot plasma, indicating that an extra highly dispersed plasma layer(s), which is most likely dominated by the intergalactic medium.

Employing $\rm|\Delta RM/\Delta DM|$, we can estimate that the magnetic field parallel to the LOS is $\langle B_{||}\rangle\approx 44\,\rm\mu G$ \citep{2023MNRAS.526.3652K}.
This magnetic field is too small to create a Faraday conversion for a cold plasma.
By synthesizing the polarimetric data of FRB 20180301A reported by \cite{2020Natur.586..693L} (Burst No. 5) and \cite{2023MNRAS.526.3652K} (B11), we estimate an average line-of-sight magnetic field strength of $\langle B_\|\rangle\sim4.6\,\rm mG$.
The average magnetic field used here is the mean value derived from the RM and DM after a long period of evolution (on the order of years).
However, considering that the circumsource medium around FRB 20180301A may exhibit strong turbulence, the magnetic field could undergo rapid changes on much shorter timescales.
RM is predominantly affected by the mixing Faraday layer, but there needs to be a mechanism that enables a reversal of the $B$-field (see Section \ref{sec5.1}).

The cold and hot plasmas give different polarized incoming waves.
The modeled circular polarization fraction of the incoming wave is $\sim12\%$ for the cold plasma and $\sim-95\%$ for the hot plasma while such a highly circular polarization degree has not been seen.
Both models show that the total polarization fraction is $\sim100\%$.
The circular polarizations generated from the intrinsic mechanism are unstable, even for a wide range of values.

\subsection{FRB 20201124A}\label{sec4.2}

\begin{figure}[ht!]
\plotone{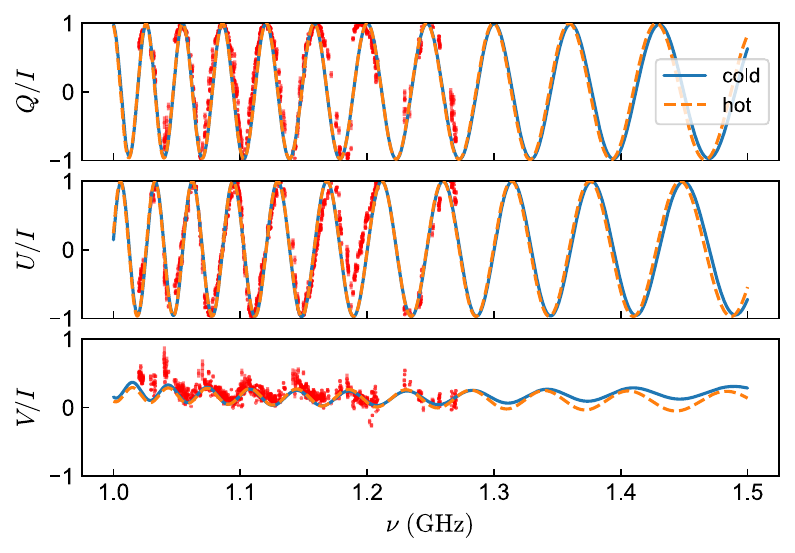}
\caption{Same as Figure \ref{fig:GFR} but for the observation data of FRB 20201124A.}
\label{fig:GFR1124}
\end{figure}

\begin{figure*}[ht!]
\plotone{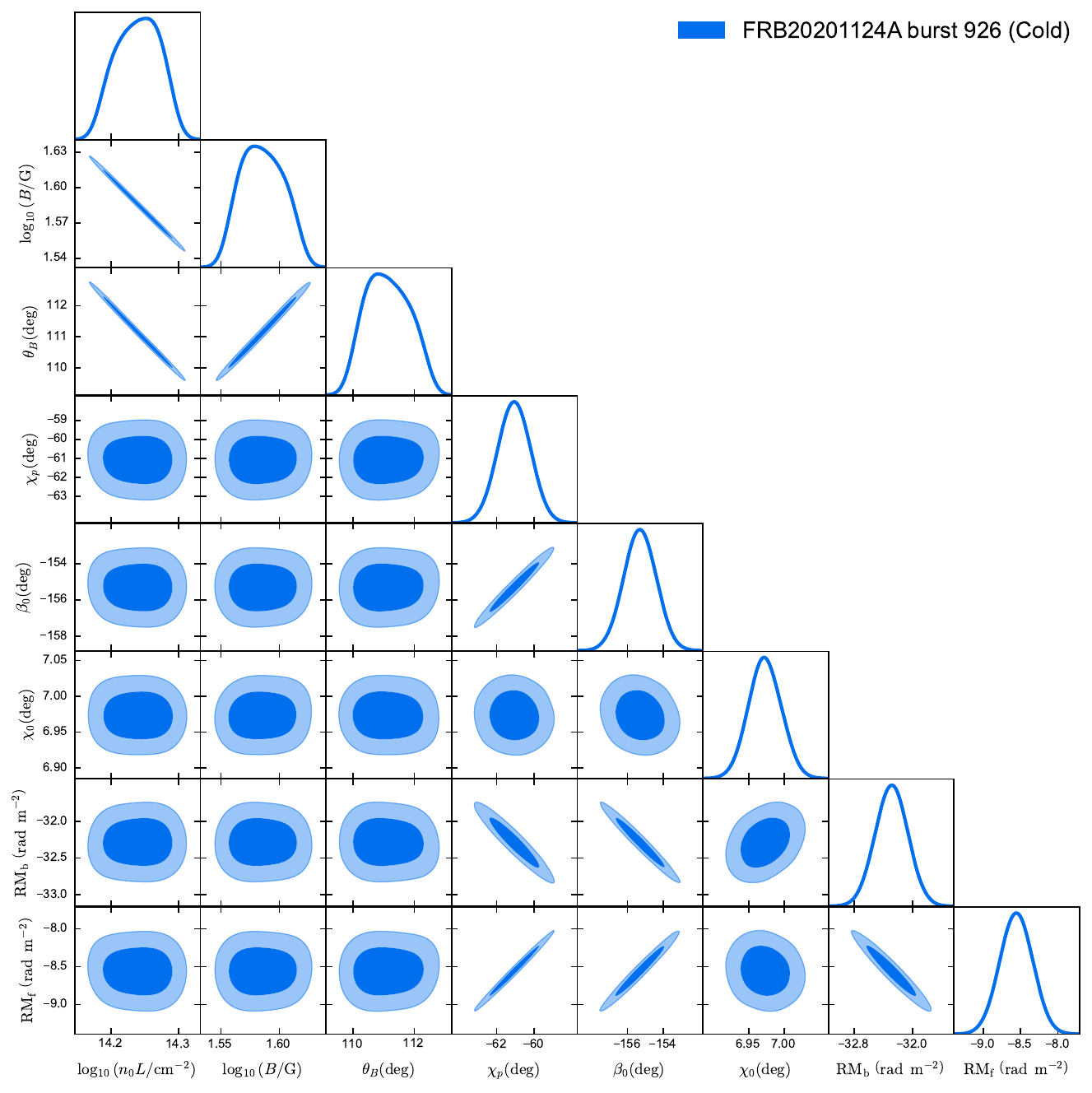}
\caption{Same as Figure \ref{fig:cold} but for FRB 20201124A.}
\label{fig:cold1124}
\end{figure*}

\begin{figure*}[ht!]
\plotone{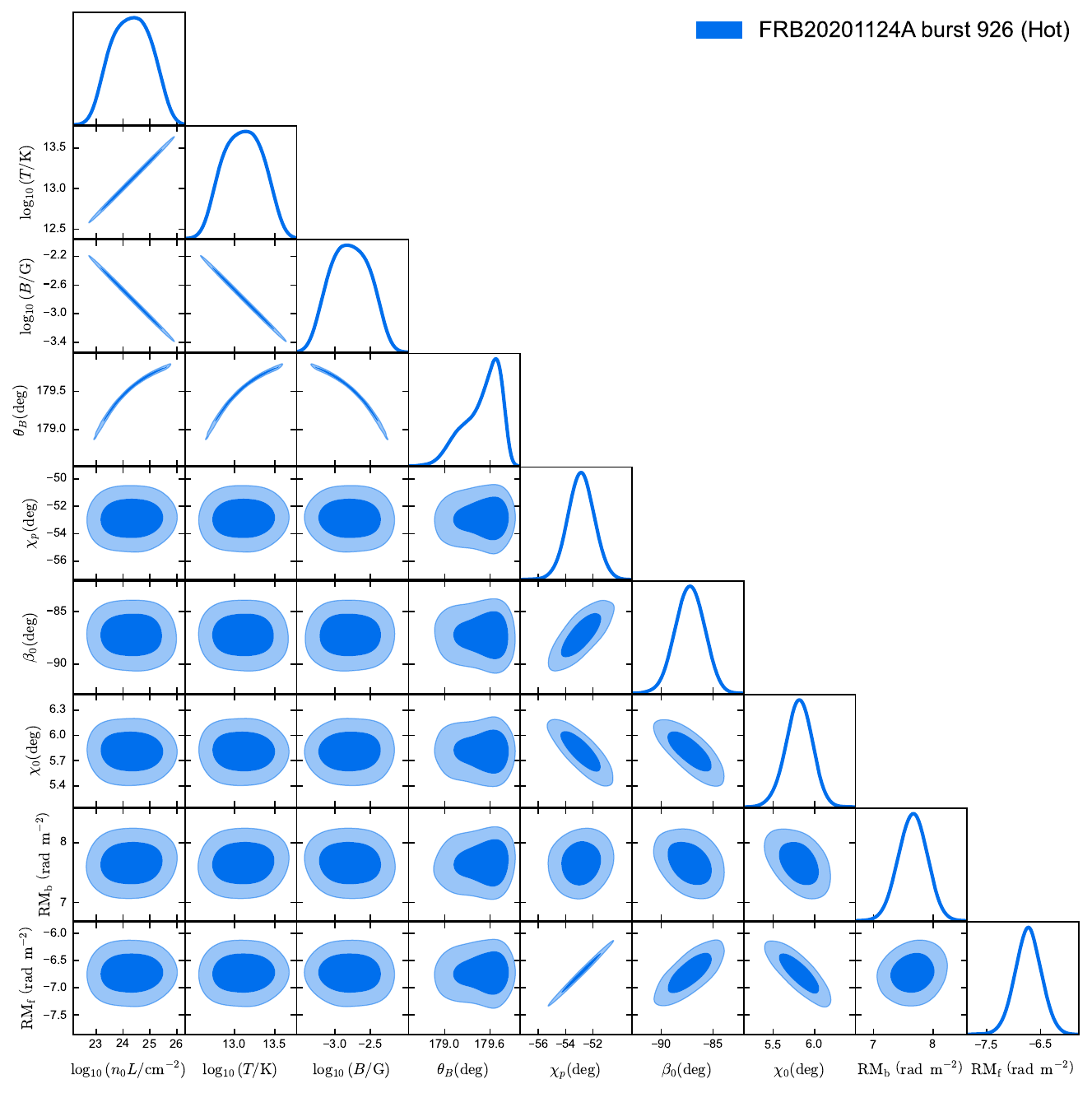}
\caption{Same as Figure \ref{fig:hot} but for FRB 20201124A.}
\label{fig:hot1124}
\end{figure*}

\begin{table*}
\centering
\caption{68.3\% credible intervals for the parameters of the cold and hot plasma model for FRB 20201124A.}
\begin{tabular}{cccccccc}
\hline\hline
 & $\log_{10}(B/\rm\,G)$ & $\theta_B$ (deg) & $\log_{10}(n_0L/\rm\,cm^{-2})$ & $\log_{10}(T/\rm \,K)$ & $\beta_0$ (deg) & $\chi_0$ (deg) & $\chi_p$ \\
Cold & $1.59\pm0.02$ & $111.1^{+0.9}_{-0.8}$ & $14.24\pm0.04$ & 0 & $-155.3\pm0.9$ & $6.97\pm0.02$ & $-61.1\pm{0.9}$\\
Hot & $-2.79^{+0.31}_{-0.30}$ & $179.6^{+0.2}_{-0.3}$ & $24.31^{+0.79}_{-0.82}$ & $13.1\pm0.3$ & $-87.2\pm{1.4}$ & $5.8\pm0.2$ & $52.9\pm1.0$\\
\hline\hline
\end{tabular}
\label{tab2}
\end{table*}

FRB 20201124A exhibits a variety of polarization features.
The burst 926 among \cite{2022Natur.611E..12X} has frequency-dependent circular polarization variation, i.e., oscillations in fractional linear and circular polarizations, as well as PA as a function of wavelength.
By using the methods in which absorptions are neglected same as FRB 20180301A, we fit the polarization spectra of the burst 926 of FRB 20201124A.
The fitted parameters are listed in Table \ref{tab2}, while Figure \ref{fig:cold1124} and Figure \ref{fig:hot1124} illustrate the marginalized posterior densities.
The modeled circular polarization fraction of the incoming wave is $\sim24\%$ for the cold plasma and $\sim20\%$ for the hot plasma. 
The foreground RMs for the cold and hot plasma mediums are ${\rm RM}_f^{\rm cold}=-8.6\pm0.2\,\rm{rad\,m^{-2}}$ and ${\rm RM}_f^{\rm hot}=-6.7\pm0.2\,\rm{rad\,m^{-2}}$, and the background RMs for them are ${\rm RM}_b^{\rm cold}=32.3\pm0.2\,\rm{rad\,m^{-2}}$ and ${\rm RM}_b^{\rm hot}=7.7\pm0.2\,\rm{rad\,m^{-2}}$.

The observed RM of FRB 20201124A ranges from $-800\,\rm rad\,m^{-2}$ to $-400\,\rm rad\,m^{-2}$, in which the contribution in the Milky Way is referred to $-51\rm\,rad\,m^{-2}$ along the direction of FRB 20201124A and that of the source rest frame is from $\rm -380\,rad\,m^{-2}$ to $-1010\rm\,rad \,m^{-2}$ \citep{2022Natur.611E..12X}.
The modeled DM of the cold plasma is $5.6\times10^{-5}\,\rm pc\,cm^{-3}$, and that of the hot plasma is $154.3\,\rm pc\,cm^{-3}$.
Both the modeled DMs are consistent with
the extragalactic DM contributions of $183-243\,\rm pc\, cm^{-3}$.
The scenario that a hot plasma cloud moves across the LOS can be dismissed, as such an occurrence would precipitate an instantaneous enhancement in DM.
Unlike FRB 20180301A, there is currently no observational evidence of RM reversal for FRB 20201124A. However, its rapidly varying RM may suggest the existence of a strong magnetic fields of $\langle B_\|\rangle>0.2$ mG on short timescales \citep{2022Natur.611E..12X}.

There is one burst exhibiting a 90.9\% circular polarization fraction among FRB 20201124A \citep{2024arXiv240803313J}.
The total polarization fraction is $\sim100\%$.
The appearance of high circularly polarized bursts is not a common phenomenon for FRB 20201124A, indicating that the mechanism behind its occurrence is highly coincidental.
Even such a highly circularly polarized burst is rare, it was unexpected in theory and unprecedented in observation in the case of FRBs.
Solar or Jovian radio emissions can have such high circular polarization, but for the sub-relativistic electrons, which can not emit extremely high bright temperatures like FRBs.
As shown in Figure \ref{fig:AbFR2}, a highly circular polarization fraction can appear at a narrowband.
The strong absorption of linear polarization reduces the intensity $I$ by approximately one order of magnitude, resulting in a high $V/I$.
If so, there may be a dense cloud with $n_0\sim10^9\,\rm cm^{-3}$ moving across the LOS to FRB 20201124A.
The size of the cloud should be much smaller than $10^{11}$ cm unless there would be a significant enhancement in DM.

\section{Discussion}\label{sec5}
\subsection{Power-law distributed plasma}\label{sec5.pl}

In the absence of thermal equilibrium within the plasma, the coefficients may have substantial deviations from those for thermal equilibrium.
A power-law distribution, i.e., $n(E)=CE^{-p}$, is a widely discussed non-equilibrium distribution.
The absorption coefficients of the power-law distributed plasma are given by
\citep{1965ARA&A...3..297G,1979rpa..book.....R,2016ApJ...822...34P,2021ApJ...921...17M}:
\begin{equation}
\begin{aligned}
\eta_I&=\frac{\sqrt{3}e^3}{8\pi m_e}\left(\frac{3e}{2\pi m_e^3c^5}\right)^{\frac{p}{2}}C(B\sin\theta_B)^{\frac{p+2}{2}}\\
&\times\Gamma\left(\frac{3p+2}{12}\right)\Gamma\left(\frac{3p+22}{12}\right)\nu^{-\frac{p+4}{2}},\\
\eta_Q&=\frac{3p+6}{3p+10}\eta_I,\\
\eta_V&=-\frac{e^3c^2}{8\sqrt{3}\pi}\left(\frac{3e}{2\pi m_e^3c^5}\right)^{\frac{p+1}{2}}C(B\sin\theta_B)^{\frac{p+3}{2}}\cot\theta_B\\
&\times\frac{(p+2)(p+3)}{p+1}\Gamma\left(\frac{3p+7}{12}\right)\Gamma\left(\frac{3p+11}{12}\right)\nu^{-\frac{p+5}{2}}.
\end{aligned}
\label{eq:9}
\end{equation}
The rotation and conversion coefficients are
\begin{equation}
\begin{aligned}
\rho_V&=\frac{e^3}{\pi m_e\nu^2}C(m_ec^2)^{-p}\frac{\ln\gamma_{\rm min}}{(p+1)\gamma_{\rm min}^{p+1}}B\cos\theta_B,\\
\rho_Q&=-\frac{e^3}{2\pi(p-2)m_e}C\left[\left(\frac{\nu}{\nu_{\rm min}}\right)^{\frac{p-2}{2}}-1\right]\\
&\times\left(B\sin\theta_B\right)^{\frac{p+2}{2}}\left(\frac{e}{2\pi m_e^3 c^5}\right)^{\frac{p}{2}}\nu^{-\frac{p+4}{2}},
\end{aligned}
\label{eq:rho_pl}
\end{equation}
where $\nu_{\rm min}=\nu_B\gamma_{\rm min}^2\sin\theta_B$.

For the regime of $\nu_{\rm min}\ll\nu$, the absorption of $V$ is much smaller than that of $Q$. The Faraday rotation rate is also much larger than the conversion rate for realistic parameter values.
The conversion rate has a frequency-dependent form similar to $\eta_I$ and $\eta_Q$. However, the conversion rate is much larger than the absorption coefficients.
Therefore, Faraday rotation is the leading effect in a relativistic medium when $\nu_{\rm min}\ll\nu$.
By comparing Equation (\ref{eq:rho_thermal}) and Equation (\ref{eq:rho_pl}), one can see that the rotation rate in a cold plasma is always larger than that in a relativistic plasma with the same $B$-field and number density.

\subsection{Reversal magnetic field}\label{sec5.1}

Faraday conversion is argued to happen in some special environments, such as relativistic plasmas or a reversal $B$-field along the LOS \citep{2019ApJ...876...74G,2019MNRAS.485L..78V,2022NatCo..13.4382W,2023Natur.618..484L}.
According to Equation (\ref{eq:rho_cold}), Faraday conversion requires that $\theta_B$ should not be too small for a cold plasma.
A hot plasma does not necessarily require a large $\theta_B$ to create a Faraday conversion.
Nevertheless, the RM is mainly contributed by the cold plasma component in a mixed hot and cold plasma.

In a realistic astrophysical environment, achieving a magnetic field that is strictly perpendicular to the LOS is difficult.
We consider that an FRB travels through the region of the ray path where the projection of the $B$-field on the ray path reverses sign.
The region is also called the quasi-transverse (QT) region, which is centered around $\theta_B=\pi/2$. The waves are 
nearly linearly polarized due to the QT approximation.
There would be an additional contribution of PA to correct for the emission due to the reversal sign of $\langle B_\|\rangle$ \citep{2010ApJ...725.1600M}.
The correction to the PA is due to the introduction of Faraday conversion when integrating the propagation equation.
The mixing Faraday scenario discussed in Section \ref{sec3.1} can occur in the QT region, where there is a reversal of RM.
Such an RM reversal has been reported from FRB 20180301A and FRB 20190520B \citep{2023Sci...380..599A,2023MNRAS.526.3652K}.

One possible scenario for RM reversal is a binary system with a magnetized companion.
An RM reversal and irregular RM variation can be produced at large magnetic inclinations for a strongly magnetized high-mass companion binary \citep{2022NatCo..13.4382W,2023ApJ...957....1X}.
No extremely large RM has been found in some known pulsars with high mass
companions \citep{2001MNRAS.325..979S,2023ApJ...943...57A}, except PSR B1259--63, a pulsar that experiences periodic eclipses by its Be star companion, exhibiting RM variations that fluctuate across zero \citep{1996MNRAS.279.1026J,2005MNRAS.358.1069J}.
The variations in the RM (both in sign and magnitude) over time in PSR B1259--63 are thought to be caused by the radio pulses traveling through a clumpy decretion disk surrounding its binary companion during the closest approach in their orbit.
RM sign change is also seen in a black widow binary PSR J2051--0827 since the change in magnetic field strength along the LOS \citep{2023ApJ...955...36W}.
The stellar wind from the highly magnetized and massive star companion plays a role in the cold plasma medium.
In that case, the Faraday mixing spectrum would appear associated with the RM reversal.
The occurrence of the Faraday mixing spectrum may be related to the orbital motion of the binary system.

\subsection{Persistent radio source}\label{sec5.2}

The hot plasma may be associated with persistent radio sources (PRSs).
Several PRSs have been discovered near several active FRB sources \citep{2017Natur.541...58C,2022Natur.606..873N,2024Natur.632.1014B,2024arXiv241201478B,2025arXiv250114247Z}. The luminosities of PRSs span in 2-3 orders of magnitude. 
Assuming that the PRS is powered by incoherent synchrotron radiation, the maximum specific luminosity is given by \citep{2020ApJ...895....7Y}
\begin{equation}
\begin{aligned}
&L_{\nu,\rm{max}}\simeq \frac{4\pi L^3n_{0,h}}{9e}m_ec^2\sigma_T B\\
&=1.6\times10^{30}\,{\rm erg\,s^{-1}\,Hz^{-1}}\left(\frac{L}{10^{17}\,\rm{cm}}\right)^3\left(\frac{B}{1\,\rm{G}}\right)\left(\frac{n_{0,h}}{1\,\rm{cm^{-3}}}\right),
\end{aligned}
\label{eq:LPRS}
\end{equation}
where $n_{0,h}$ only denotes the number density of the hot plasma component. One can also show that the specific luminosity is proportional to the $\rm |RM|$ of the FRB source \citep{2020ApJ...895....7Y,2022ApJ...928L..16Y}, which is observationally confirmed \citep{2024Natur.632.1014B,2024arXiv241201478B,2025arXiv250114247Z}.
However, to generate observed PRS luminosity of FRB 20201124A, the fitting results for the hot plasma lead to a plasma size of $L\sim 10^{14}$ cm, thus the number density of $n_0\sim10^{10}\,\rm cm^{-3}$.
If so, there would be strong absorption in the plasma medium and the Stokes intensities would decrease significantly.
The PRS could be a compact nebula, which itself does not meet the necessary conditions to produce Faraday conversion.

Intriguingly, FRB 20190520B is associated with a PRS, which has a flux density of $\sim200\,\rm\mu Jy$ at 3 GHz and spectral radio luminosities of the order of $10^{29}\,\rm erg\,s^{-1}\,Hz^{-1}$ \citep{2022Natur.606..873N,2023ApJ...959...89Z}.
This PRS may be a Faraday screen on the propagation path of the FRB.
Combined with the $\rm\Delta DM$ and $\rm\Delta RM$, the average magnetic field strength along the LOS of FRB 20190520B can be estimated as $3-6$ mG \citep{2023Sci...380..599A}.
Under these conditions, the Faraday conversion may be comparable to Faraday rotation, leading to a frequency-dependent circular polarization.
The polarization may be more complex due to the strong depolarization in low-frequency bands \citep{2022Sci...375.1266F}.

Observation of FRB 20190520B gives a transverse physical size constraint of $< 9$ pc, while a lower limit of $\gtrsim 0.22 $ pc can be placed
from the lack of a clear break in the synchrotron spectrum due to self-absorption \citep{2023ApJ...958L..19B,2023ApJ...959...89Z}.
We take $L\sim10^{18}\,\rm cm$.
In order to archive the observed PRS luminosity, one needs $n_{0,h}\sim0.1\,\rm cm^{-3}$ with magnetic field strength of $\sim1$ mG.
This number density is only for the relativistic electrons.
If the medium contains $10^4$ times more non-relativistic electrons than relativistic ones, one can estimate a DM of $\sim 10^3\rm\,pc\,cm^{-3}$, matching the referred high DM of FRB 20190520B \citep{2022Natur.606..873N}.
Under such conditions, RM is mainly contributed by the dense cold plasma component.
The hot plasma component may contribute to a larger Faraday conversion coefficient, while it is much smaller than the Faraday rotation coefficient of the cold plasma.

The high DM and RM of FRB 20190520B hint toward a
dense and highly magnetized circum-source medium.
The EVN limit of $<9$ pc for a nebula's size with the expanding supernova ejecta shell gives an age of at least 900 years by assuming an ejecta velocity of $10^4\,\rm km\,s^{-1}$ \citep{2023ApJ...958L..19B}.
We can estimate the number density to $n_0\lesssim3.9(M_{\rm ej}/10M_\sun)\,\rm cm^{-3}$, which is not sufficient to provide the large DM, where $M_{\rm ej}$ is the ejecta mass.
A hyper-nebula powered by a central engine accreting at $\sim10^6$ Eddington-limited mass transfer rate for a black hole with a mass of $10M_\sun$, is proposed to explain the PRS \citep{2022ApJ...937....5S}.
The strong and rapid RM variation, including possible sign reversals, is attributed to turbulent motions in the nebula.

Alternatively, an intermediate-mass black hole may account for the PRS.
The massive black hole model can interpret large variations of DM and RM for both FRB 20121102A and FRB 20190520B (e.g., \citealt{2018ApJ...852L..12D}), even for a ``zero-crossing'' reversal.
However, no X-ray counterpart of the PRS has been observed, which gives an upper limit of isotropic X-ray luminosity of $L_X<5\times10^{37}\,\rm erg\,s^{-1}$ for FRB 20121102A \citep{2017Natur.541...58C}.
A relationship between black hole mass and luminosity of radio ($L_R$) and X-rays is obtained as $\quad \log \left(M / 10^8 M_{\odot}\right)=0.55+1.09 \log \left(L_R / 10^{38} \mathrm{erg}\,\mathrm{s}^{-1}\right)-$ $0.59 \log \left(L_X / 10^{40} \mathrm{erg}\, \mathrm{s}^{-1}\right)$ based on a sample of 30
AGN with independent dynamical mass measurements \citep{2019ApJ...871...80G}.
One can estimate a black hole mass of $\sim10^{11}M_\sun$ much larger than the stellar mass of the host galaxy of FRB 20121102A.
The host galaxy of FRB 20190520B has an estimated stellar mass of $\sim6\times10^8M_\sun$, suggesting the presence of an intermediate-mass black hole with a mass around $\sim10^{2-6}M_\sun$.
The mass estimation implies a modeled X-ray luminosity lower limit of $2.1\times10^{44}\,\rm erg\,s^{-1}$ that is consistent with the Fermi-LAT observation \citep{2024arXiv240212084Y}.

In contrast, FRB 20180301A has not been discovered to be accompanied by a PRS \citep{Bhandari_2022}.
Combined with $\langle B_{||}\rangle\sim4.6$ mG, the number density of relativistic plasma is $n_{0,h}\lesssim10^{-2}\,{\rm cm^{-3}}(L/10^{17}\,\rm cm)^{-3}$, much smaller than the fitting results of the hot plasma medium in Section \ref{sec4.1}.
Considering a strong turbulent environment, the magnetic field perpendicular to the LOS could be extremely high in a short time.
This suggests that the environment of FRB 20180301A may be situated within a binary system.
The stellar atmosphere or the stellar wind of a compact star could effectively explain the variations in DM and RM, including the sign reversal of RM.

\subsection{Other FRB sources and pulsars}\label{sec5.4}

The small and non-variable RMs of FRB 20220912A may be attributed to a clean environment.
Highly circularly polarized bursts from FRB 20220912A are likely to be due to intrinsic radiation mechanisms rather than propagation effects \citep[e.g.][]{2022MNRAS.517.5080W,2023MNRAS.522.2448Q}.
Circular polarization in some bursts varies with frequency but without an obvious oscillation frequency \citep{2023ApJ...955..142Z}. 
The complex spectro-polarimetric properties require special magnetic field environments, e.g., in a binary star system that is not
consistent with a clean environment.

XTE J1810--197 can also display linear-to-circular conversion from radio emissions after its 2018 outburst \citep{2024NatAs...8..606L}.
The rotation of the polarization vector is different from standard Faraday rotation or conversion.
Other propagation effects, such as
coherent/partially-coherent mode mixing, cannot explain the well-defined frequency dependence \citep{2023MNRAS.525..840O}.
However, no well-defined phase differences between $Q$, $U$, and $V$ can be measured, and the total polarization is frequency-dependent.
These phenomena suggest strong absorption and depolarization of the Faraday screen.
In addition, two pulsars, PSR J0835--4510 and J1644--4559, have slightly frequency-dependent circular polarization \citep{2024MNRAS.534.2485U}, while it is hard to distinguish whether it is due to an intrinsic origin or a propagation effect.

\section{Summary}\label{sec6}

Rich polarization data from FRBs have been observed, showing a complicated frequency evolution of Stokes parameters in some cases. We developed an analytical solution for the transformation of the polarized emission on the Stokes basis.
We mainly focus on the spectro-polarimetric properties of an outgoing wave from a homogeneous magnetized thermal plasma medium.
We have drawn the following conclusions.

(1) When an FRB propagates through a magnetized thermal plasma, the emissivity of the medium can be neglected. The absorption and Faraday mixing coefficients mainly influence the polarization spectrum of the outgoing waves.

(2) For a low-density cold thermal plasma medium, Faraday rotation can dominate the propagation for GHz-waves if $B\lesssim3\times10^2$ G, while Faraday conversion becomes strong
for $B\gtrsim3\times10^2$ G.
Absorption becomes dominant for a highly dense medium.
In a hot thermal plasma with $\nu_B\ll\nu$, the Faraday rotation rate is significantly reduced, while the conversion rate increases compared to a cold plasma.
Furthermore, Faraday rotation is the only dominant process for a power-law distributed relativistic plasma when $\nu_{\rm min}\ll\nu$.

(3) Spectro-polarimetric properties are diverse for a Faraday mixing scenario, i.e., $|\rho_V|\approx|\rho_Q|$.
Stokes parameters, $Q$, $U$, and $V$, oscillate with wavelength but for different phases, depending on the direction of the $B$-field projection on the LOS. Linear polarization have approximately
twice the oscillation frequency as circular polarization.
The rotational axis of the polarization spectrum undergoes precession as the frequency changes due to the difference of frequency dependencies between Faraday rotation and conversion.

(4) The PA spectrum within one period can sometimes be characterized as an `S-shape' or other more complex shapes. 
In this case, the PA spectrum deviates from a $\lambda^2$-oscillation, reminiscent of what is known as the GFR.
We show that the Faraday mixing case can replace the method of GFR to study polarization spectra, which can give a more physical representation of the problem, as the Faraday mixing spectrum reflects the magnetic environment more clearly.

(5) Such a Faraday mixing scenario can happen for both the cold and hot thermal plasmas.
In that case, the cold plasma medium requires a strong magnetic field component perpendicular to the LOS.
For the hot plasma medium, the strength of the magnetic field and the angle it makes relative to the LOS could be much smaller.
To display a similar polarization spectrum, the hot plasma medium should have a larger DM than the cold plasma.

(6) Significant absorption combined with Faraday rotation and conversion can exist in a highly dense and magnetized plasma medium.
The outgoing waves can be highly circularly polarized in a narrow emitting band when the incoming waves have a certain degree of circular polarization.

(7) We apply the transformation equation to study the magnetic environments of some FRBs.
The frequency-dependent circular polarization properties of FRB 20180301A and FRB 20201124A can be interpreted through the Faraday mixing scenario.
The circum-source medium may be predominantly influenced by the binary companion that experiences significant turbulence, which can lead to a reversal of the RM, thereby inducing Faraday conversion.
Some dense clouds may move across the LOS of FRB 20201124A, which induces highly circularly polarized bursts due to absorption.
In the case of FRB 20190520B, the circum-source medium may consist of a combination of cold and hot plasmas, with the hot plasma emitting as the PRS and the cold plasma contributing to the high DM.
Bursts with frequency-dependent circular polarizations are expected to be observed in FRB 20240114A.

\section*{Acknowledgments}
We are grateful to the referee for constructive comments,
and Shunshun Cao, Alex Cooper, Ruxi Liang, Zenan Liu, Rui Luo, Jiarui Niu, Lucy Oswald, Yuanhong Qu, Pavan Uttarkar, Pei Wang, Siyue Wang, Tiancong Wang, Yuan-Pei Yang, Junshuo Zhang, Dejaing Zhou, and Yuanchuan Zou for helpful discussion.
W.-Y. W. acknowledges support from the NSFC (No.12261141690 and No.12403058), the National SKA Program of China (No. 2020SKA0120100), and the Strategic Priority Research Program of the CAS (No. XDB0550300).
C.-H. N. is supported by NSFC (No.12203069), the CAS Youth Interdisciplinary Team, and the Foundation of Guizhou Provincial Education Department for Grants No. KY(2023)059.
J.F.L. acknowledges support from the NSFC (Nos.11988101 and 11933004) and from the New Cornerstone Science Foundation through the New Cornerstone Investigator Program and the XPLORER PRIZE.
W.W.Z. is supported by the National Natural Science Foundation of China (grant No. 12041303) and National SKA Program of China (grant No. 2020SKA0120200)

\appendix
\section{Plasma Response}\label{response}

A general geometry is depicted in Figure \ref{fig:e1e2e3}.
PAs increase as the RCP wave propagates in the positive $k$ direction, that is, the electric field vector rotates counter-clockwise, as seen by the observer.
Such an RCP wave defined here forms a left-handed helix in space.
It is worth noting that when the definitions of LCP and RCP differ, the direction of Faraday rotation and conversion may also be reversed.
In this work, all our calculations follow the PSR/IEEE convention \citep{2010PASA...27..104V}.
With this convention, the signs of some parameters differ from \cite{2008ApJ...688..695S} and \cite{2011MNRAS.416.2574H}.

The response tensor can describe the propagation of weak electromagnetic waves in a homogeneous magnetized plasma.
The dielectric tensor is given by
\begin{equation}
\epsilon^\mu_\nu=\delta^\mu_\nu+\frac{4\pi c}{\omega^2}\alpha^\mu_\nu,
\end{equation}
where the response tensor is given by \citep{1958PhDT........18T}
\begin{equation}
\begin{gathered}
\alpha_\nu^\mu(k)=\frac{i e^2 n_0 \omega}{c m \bar\gamma^{2}K_2(\bar\gamma^{-1})} \int_0^{\infty} d \xi\left[t_\nu^\mu \frac{K_2(r)}{r^2}-R^\mu \bar{R}_\nu \frac{K_3(r)}{r^3}\right], \\
t_\nu^\mu=\left(\begin{array}{cc}
\cos ^2 \theta_B \cos \omega_B \xi+\sin ^2 \theta_B & \eta \cos \theta_B \sin \omega_B \xi \\
-\eta \cos \theta_B \sin \omega_B \xi & \cos \omega_B \xi
\end{array}\right), \\
R^\mu=\frac{\omega \sin \theta_B}{\omega_B}\left[\cos \theta_B\left(\sin \omega_B \xi-\omega_B \xi\right),-\eta\left(1-\cos \omega_B \xi\right)\right] \\
\bar{R}_\nu=\frac{\omega \sin \theta_B}{\omega_B}\left[\cos \theta_B\left(\sin \omega_B \xi-\omega_B \xi\right), \eta\left(1-\cos \omega_B \xi\right)\right] \\
r=\left[\frac{1}{\bar\gamma^2}-2 i \frac{\omega \xi}{\bar\gamma} +\frac{\omega^2 \sin ^2 \theta_B}{\omega_B^2}\left(2-\omega_B^2 \xi^2-2 \cos \omega_B \xi\right)\right]^{1 / 2},
\end{gathered}
\label{eq:response}
\end{equation}
where $\eta$ is the sign of the charge.
Note that the only significant contribution to the integral over proper time is from the range $\omega_B\xi\ll1$.
The first terms of Equation (\ref{eq:response}) can be written as
\begin{equation}
\begin{gathered}
r^2=r_0^2+\delta r^2, \quad r_0^2=\frac{1}{\bar\gamma^2}-2 i \frac{\omega \xi}{\bar\gamma}, \quad \delta r^2=-\frac{\sin ^2 \theta_B}{12} \omega^2 \omega_B^2\xi^4, \\
t_\nu^\mu=\left(\begin{array}{cc}
1-\cos ^2 \theta_B\left(\omega_B^2\xi^2 / 2\right) & \eta\omega_B\xi \cos \theta_B \\
-\eta\omega_B\xi \cos \theta_B & 1-\omega_B^2 \xi^2 / 2
\end{array}\right), \\
R^\mu \bar{R}_\nu=-\frac{\omega^2 \omega_B^2\xi^4}{4} \sin ^2 \theta_B\left(\begin{array}{cc}
0 & 0 \\
0 & 1
\end{array}\right) .
\end{gathered}
\end{equation}
The components of the dielectric tensor
in the lowest orders in $\omega_B/\omega$ are given by \citep{2008ApJ...688..695S,2011MNRAS.416.2574H}
\begin{equation}
\begin{aligned}
\varepsilon_1^1&=1-\frac{\omega_p^2}{\omega^2}\left[\frac{K_1(\bar\gamma^{-1})}{K_2(\bar\gamma^{-1})}\left(1+\frac{\omega_B^2}{\omega^2} \cos ^2 \theta_B\right)+\frac{\bar\gamma\omega_B^2 \sin ^2 \theta_B}{\omega^2}\right] \\
\varepsilon_2^2&=1-\frac{\omega_p^2}{\omega^2}\left[\frac{K_1(\bar\gamma^{-1})}{K_2(\bar\gamma^{-1})}\left(1+\frac{\omega_B^2}{\omega^2}\right)+\frac{7\bar\gamma \omega_B^2 \sin ^2 \theta_B}{\omega^2}\right] \\
\varepsilon_2^1&=-\varepsilon_1^2=-i \eta \frac{\omega_p^2 \omega_B}{\omega^3} \frac{K_0(\bar\gamma^{-1})}{K_2(\bar\gamma^{-1})} \cos \theta_B
\end{aligned}
\end{equation}
where $\omega_p$ is the plasma frequency.
The Faraday rotation and conversion coefficients in the bias shown in Figure \ref{fig:e1e2e3} are defined as
\begin{equation}
\begin{aligned}
\rho_V&=i\frac{\omega}{c}\epsilon^1_2,\\
\rho_Q&=-\frac{\omega}{2c}(\epsilon^1_1-\epsilon^2_2).
\end{aligned}
\end{equation}
In order to consider more general cases, 
two multipliers could be introduced to fix the response tensor.
These two multipliers contain information about higher orders.
One expression of the two multipliers is given by \citep{2008ApJ...688..695S}
\begin{equation}
\begin{aligned}
f(X)&=2.011 \exp \left(-\frac{X^{1.035}}{4.7}\right) -\cos \left(\frac{X}{2}\right) \exp \left(-\frac{X^{1.2}}{2.73}\right)-0.011 \exp \left(-\frac{X}{47.2}\right),\\
g(X)&=1-0.11\ln{(1+0.035X)},\\
X&=10^{3/2}2^{1/4}\bar\gamma\left(\frac{\omega_B}{\omega}\sin\theta_B\right)^{1/2}.
\end{aligned}
\end{equation}
The boundary of the expression is that when the high-frequency limit ($\omega\gg\omega_p$) is valid.
For a cold plasma medium with $\omega_B\ll\omega$, the coefficients can be calculated as
\begin{equation}
\begin{aligned}
&\rho_V=\frac{e^3Bn_0\cos\theta_B}{\pi m_e^2c^2\nu^2},\\
&\rho_Q=-\frac{e^4B^2n_0\sin^2\theta_B}{4\pi^2c^3m_e^3\nu^3};
\end{aligned}
\label{eq:rho_cold}
\end{equation}
while for a hot plasma, the coefficients can be calculated as
\begin{equation}
\begin{aligned}
&\rho_V=\frac{e^3Bn_0\cos\theta_B}{2\pi k_B^2T^2\nu^2}\ln{\frac{k_BT}{m_ec^2}},\\
&\rho_Q=-\frac{3e^4B^2n_0k_BT\sin^2\theta_B}{2\pi^2c^4m_e^4\nu^3}.
\end{aligned}
\label{eq:rho_hot}
\end{equation}
As can be seen, the formations of Faraday conversion and rotation coefficients have the same sign for electrons.
If the charges are positrons, the signs of these two coefficients can be opposite.

\section{Generalized Faraday rotation}\label{apendix:GFR}
In addition to the conventional Faraday rotation, the GFR introduces a scaling index $\alpha$, where $\alpha$ is no longer frozen to 2, to describe the spectral dependence.
scaling index $\alpha$. 
The polarization vector is defined as
\begin{equation}
\mathbf{P}(\lambda)=P\left(\begin{array}{c}
\cos (2 \Psi) \cos (2 \chi) \\
\sin (2 \Psi) \cos (2 \chi) \\
\sin (2 \chi)
\end{array}\right) \text {.}
\label{eq:Pvector}
\end{equation}
Here $\Psi$ is expressed as
\begin{equation}
\Psi(\lambda)=\Psi_0+\operatorname{GRM}\left(\lambda^\alpha-\lambda_0^\alpha\right),
\end{equation}
where $\Psi_0$ is the intrinsic PA,  $\chi$ is the ellipticity angle with respect to the $V$-axis in the Poincar\'e sphere, and GRM is an analog of RM for the conventional Faraday rotation.
The linear-to-circular conversion is established by applying a pair of rotation matrices to shift the longitude and latitude of the polarization-rotation axis.
The final polarization vector can be written as \citep{2024NatAs...8..606L}
\begin{equation}
\mathbf{P}_{\mathrm{GFR}}=\mathbf{R} \psi \mathbf{R}_{\theta \phi} \mathbf{P}(\lambda),
\end{equation}
where $\mathbf{R}_\psi$ is the FR-induced shift in the polarization vector, with a fixed latitude on the Poincare sphere, $\mathbf{R}_{\theta \phi}$ is the rotation matrix describing the arbitrary rotation of the polarization vector on Poincare sphere over different latitudes:
\begin{equation}
\begin{aligned}
& \mathbf{R}_\psi=\left(\begin{array}{ccc}
\cos 2 \psi & -\sin 2 \psi & 0 \\
\sin 2 \psi & \cos 2 \psi & 0 \\
0 & 0 & 1
\end{array}\right), \\
& \mathbf{R}_{\theta \phi}=\left(\begin{array}{ccc}
\cos (\theta) \cos (\phi) & -\cos (\theta) \sin (\phi) & \sin (\theta) \\
\sin (\phi) & \cos (\phi) & 0 \\
-\sin (\theta) \cos (\phi) & \sin (\theta) \sin (\phi) & \cos (\theta)
\end{array}\right),
\end{aligned}
\end{equation}
where $\theta$ describes the angle of the axis about which the polarization vector rotates, and $\phi$ is the rotation about the Stokes-$U$ axis on the Poincare sphere.

\bibliography{sample631}{}
\bibliographystyle{aasjournal}



\end{document}